\pdfoutput=1

\documentclass[iop, apj, twocolappendix, numberedappendix]{emulateapj}

\usepackage{xspace}
\usepackage{amsmath}
\usepackage{framed} 
\usepackage{txfonts}
\usepackage{epstopdf} 
\usepackage{color}
\usepackage{rotating}
\usepackage{natbib}
\usepackage{ulem}
\usepackage{xspace}
\usepackage[colorlinks=true,urlcolor=blue,linkcolor=blue,citecolor=blue]{hyperref}

\newcommand{\kpch}{\>{h^{-1}{\rm kpc}}}

\newcommand{\msun}{\>{M_{\odot}}}

\newcommand{\msunh}{\>h^{-1} M_\odot}

\newcommand{\kms}{\>{\rm km}\,{\rm s}^{-1}}

\def\mvir{M_{\rm vir}}
\def\rvir{R_{\rm vir}}
\def\gcm3{\mathrm{g} / \mathrm{cm}^3}
\def\m200m{M_{\rm 200m}}

\def\gtsima{$\; \buildrel > \over \sim \;$}
\def\ltsima{$\; \buildrel < \over \sim \;$}
\def\prosima{$\; \buildrel \propto \over \sim \;$}
\def\gsim{\lower.7ex\hbox{\gtsima}}
\def\lsim{\lower.7ex\hbox{\ltsima}}
\def\simgt{\lower.7ex\hbox{\gtsima}}
\def\simlt{\lower.7ex\hbox{\ltsima}}
\def\simpr{\lower.7ex\hbox{\prosima}}

\def\rmc{{\rm c}}
\def\rmd{{\rm d}}
\def\rme{{\rm e}}
\def\rmf{{\rm f}}

\def\rmi{{\rm i}}

\def\rmm{{\rm m}}

\def\rms{{\rm s}}
\def\rmt{{\rm t}}

\def\rs{r_{\rm s}}
\def\mstar{M_{\ast}}

\def\rdelta{R_{\Delta}}
\def\mdelta{M_{\Delta}}

\def\rsp{R_{\rm sp}}
\def\msp{M_{\rm sp}}

\def\mfrs{M_{<4r_\rms}}

\def\rinfall{R_{\rm infall}}
\def\minfall{M_{\rm infall}}

\def\fpe{f_{\rm bdry}}

\def\rhoc{\rho_{\rm c}}

\def\rhoref{\rho_{\rm ref}}

\def\cvir{c_{\rm vir}}

\def\mtom{M_{\rm 200m}}
\def\rtom{R_{\rm 200m}}

\def\mtoc{M_{\rm 200c}}
\def\rtoc{R_{\rm 200c}}

\def\mfoc{M_{\rm 500c}}

\def\rtfc{R_{\rm 2500c}}

\def\vpeak{V_{\rm peak}}

\usepackage{etoolbox}
\makeatletter
\patchcmd{\NAT@citex}
  {\@citea\NAT@hyper@{\NAT@nmfmt{\NAT@nm}\NAT@date}}
  {\@citea\NAT@nmfmt{\NAT@nm}\NAT@hyper@{\NAT@date}}
  {}
  {}
\patchcmd{\NAT@citex}
  {\@citea\NAT@hyper@{%
     \NAT@nmfmt{\NAT@nm}%
     \hyper@natlinkbreak{\NAT@aysep\NAT@spacechar}{\@citeb\@extra@b@citeb}%
     \NAT@date}}
  {\@citea\NAT@nmfmt{\NAT@nm}%
   \NAT@aysep\NAT@spacechar%
   \NAT@hyper@{\NAT@date}}
  {}
  {}
\patchcmd{\NAT@citex}
  {\@citea\NAT@hyper@{%
     \NAT@nmfmt{\NAT@nm}%
     \hyper@natlinkbreak{\NAT@spacechar\NAT@@open\if*#1*\else#1\NAT@spacechar\fi}%
       {\@citeb\@extra@b@citeb}%
     \NAT@date}}
  {\@citea\NAT@nmfmt{\NAT@nm}%
   \NAT@spacechar\NAT@@open\if*#1*\else#1\NAT@spacechar\fi%
   \NAT@hyper@{\NAT@date}}
  {}
  {}
\makeatother

\shorttitle{The Splashback Radius}
\shortauthors{More, Diemer \& Kravtsov}

\slugcomment{To be submitted to the Astrophysical Journal}

\begin{document}


\def\figdir{.}
\def\figext{pdf}


\title{The splashback radius as a physical halo boundary and the growth of halo mass}
\author{Surhud More \altaffilmark{1}, Benedikt Diemer \altaffilmark{2,3}, and Andrey V. Kravtsov\altaffilmark{2,3,4}}

\affil{
$^1$ Kavli Institute for the Physics and Mathematics of the Universe (WPI), The
University of Tokyo,\\ 5-1-5 Kashiwanoha, Kashiwa-shi, Chiba, 277-8583, Japan;{\tt surhud.more@ipmu.jp}\\
$^2$ Department of Astronomy and Astrophysics, The University of Chicago, Chicago, IL 60637 USA \\
$^3$ Kavli Institute for Cosmological Physics, The University of Chicago, Chicago, IL 60637 USA \\
$^4$ Enrico Fermi Institute, The University of Chicago, Chicago, IL 60637 USA \\
}


\begin{abstract}
The boundaries of cold dark matter halos are commonly defined to enclose a density contrast $\Delta$ relative to a reference (mean or critical) density. We argue that a more physical halo boundary choice is the radius at which accreted matter reaches its first orbital apocenter after turnaround. This {\it splashback radius}, $\rsp$, manifests itself as a sharp density drop in the halo outskirts, at a location that depends upon the mass accretion rate. We present calibrations of $\rsp$ and the enclosed mass, $\msp$, as a function of mass accretion rate and peak height. We find that $\rsp$ is in the range $\approx0.8-1\rtom$ for rapidly accreting halos and is $\approx1.5\rtom$ for slowly accreting halos. Thus, halos and their environmental effects can extend well beyond the conventionally defined ``virial'' radius. We show that $\msp$ and $\rsp$ evolve relatively strongly compared to other commonly used definitions. In particular, $\msp$ evolves significantly even for the smallest dwarf-sized halos at $z=0$. We also contrast $\msp$ with the mass enclosed within four scale radii of the halo density profile, $\mfrs$, which characterizes the inner halo. During the early stages of halo assembly, $\msp$ and $\mfrs$ evolve similarly, but in the late stages $\mfrs$ stops increasing while $\msp$ continues to grow significantly. This illustrates that halos at low $z$ can have ``quiet'' interiors while continuing to accrete mass in their outskirts. We discuss potential observational estimates of the splashback radius and show that it may already have been detected in galaxy clusters.
\end{abstract}



\section{Introduction}
\label{sec:intro}

In the standard paradigm for structure formation, galaxies form through the dissipative condensation of baryons at the centers of bound clumps of dark matter, called halos, that form within nodes of the cosmic web \citep{rees_77, white_78, fall_80, blumenthal_84}. The growth rate of  halos controls the rate at which baryons are accreted, and complex astrophysical processes determine how efficiently these baryons are converted to stars \citep[see, e.g.,][]{kauffmann_93}. Therefore, scaling relations between the luminosity or stellar mass of galaxies and the mass of their dark matter halos can provide physical insights into the process of galaxy formation \citep[see, e.g.,][]{kravtsov_04,tasitsiomi_04,vale_04, conroy_05_sk, mandelbaum_06_shmr, conroy_09_sham, more_09_sk, more_11_sk, more_14, cacciato_09, cacciato_13, moster_10_shmr, behroozi_10_shmr, behroozi_13_shmr, zehavi_11, leauthaud_12_shmr, kravtsov_13, mccracken_etal14, kravtsov_14}. Such inference, however, requires a basic understanding of the rate of growth of both the halos and stellar components of galaxies.

The halo boundary within which halo mass is measured is usually defined as a radius, $\rdelta$, of a sphere enclosing a certain density contrast $\Delta$ with respect to a chosen reference density $\rhoref$,
\begin{equation}
\mdelta = \frac{4}{3}\pi \rdelta^3 \Delta \rhoref\,.
\end{equation}
Various values of $\Delta$ are commonly used, some motivated by the analytic solution for the ``virialization overdensity'' in models of the collapse of a top-hat spherical density perturbation, others motivated by the extent to which observations can reliably measure masses (such as $\rtfc$). Common choices for the reference density include the mean matter and critical densities of the universe, both of which evolve with redshift.\footnote{Although in the case of $\Lambda$CDM model, the evolution of $\rhoref = \rhoc(z)$ saturates at $z < 0$ as the energy density starts to be dominated by the cosmological constant.}  

However, the collapse of realistic density peaks in CDM models is considerably more complex than is envisioned in the spherical top-hat collapse model. First, the peaks are not spherical and their collapse rate depends on tidal forces which, in turn, depend on the shape of the peak \citep[e.g.,][]{bond_myers96, dalal_08}. Second, although the density within the peaks is nearly flat at the center (i.e., resembles a top-hat profile), it systematically decreases (on average) with increasing distance from the peak center \citep[e.g.,][]{bardeen_etal86, dalal_10}. The collapse of such peaks is thus extended in time and is not characterized by a well-defined ``virialization'' epoch. Moreover, the successive collapse of density shells results in an extended and smooth density profile. Finally, the peaks have substructure on smaller scales, which results in the collapse and merging of smaller peaks during the collapse of a given global peak. Such mergers are accompanied by non-linear interactions and the redistribution of mass from small to large radii \citep{valluri_07}. In particular, in major mergers a significant fraction of the progenitor material ends up at radii beyond the commonly used halo boundary \citep{kazantzidis_etal06}, meaning that mass is not additive in halo mergers when standard mass definitions are used. 

A further complication is that $\rdelta$ for a given choice of $\Delta$  at some epoch $z_1$ may not be sufficiently large to enclose all of the mass accreted by a halo prior to that epoch. In this case, the change of mass between $z_1$ and some later epoch $z_2 < z_1$ will include both the new mass accreted between $z_1$ and $z_2$ and the mass that was accreted at $z > z_1$, but was located at $r > \rdelta(z_1)$. 
Thus, in the absence of any actual accretion, $\rdelta$ grows due to the decreasing reference density, but the resulting change in mass cannot be interpreted as physical mass growth, change in the halo potential, or as an indication of ongoing accretion. 

A number of recent studies \citep[see, e.g.,][]{prada_06, diemand_vl_2007, cuesta_2008, diemer_13_pe, zemp_14, diemer_14_pro} have argued that, on average, halos of mass $\lesssim 10^{12} \msunh$ accrete little new mass at low $z$ at radii $\lesssim \rdelta$ if $\Delta \gtrsim 200-300$ times the critical density of the universe. The increase in their mass $\mdelta$ is largely due to the change of their boundary $\rdelta$ in response to a decreasing reference density, not due to the physical accretion of new matter within $\rdelta$. In \citet{diemer_13_pe}, we quantified the amount of halo mass growth that can be attributed to such {\it pseudo-evolution}, and found that it contributes significantly to the overall mass growth of low-mass halos since $z \sim 1$. Given that standard mass definitions are subject to pseudo-evolution, it raises an important question: do galaxy-sized halos accrete new matter at late times and if so, at what rate?

In this paper, we carefully consider particular choices for the halo radius and mass definition. We argue that the most natural and physical halo boundary can be identified with the radius at which newly accreted matter is reaching its first orbital apocenter after its initial turnaround.  This {\it splashback radius} corresponds to the outer caustic in the spherical models of secondary collapse \citep{fillmore_goldreich84, bertschinger85, adhikari_14}. The splashback radius also physically separates the region where matter is infalling for the first time and the region occupied by matter that has orbited through the central halo region at least once.  The mass within the splashback radius is thus guaranteed to include all of the mass that was accreted by a given redshift $z$. As shown recently \citep[][see also \citealt{adhikari_14}]{diemer_14_pro}, the halo density profile exhibits a sharp steepening of its slope around the splashback radius, in correspondence to the density jump expected at this radius in analytic collapse models \citep{fillmore_goldreich84, bertschinger85, adhikari_14}. We compare the evolution of the splashback radius, $\rsp$, and the corresponding mass, $\msp \equiv M(<\rsp)$, with standard mass definitions such as $\mvir$, $\mtoc$, and $\mtom$. Furthermore, we contrast $\msp$ with the mass within a fixed multiple of the halo scale radius, $\mfrs$. This mass definition characterizes the evolution of the inner regions of halos and is manifestly unaffected by pseudo-evolution.

The paper is organized as follows. In Section \ref{sec:peakcollapse} we discuss analytical models of peak collapse and motivate our choice of the splashback radius as the physical boundary of a halo. In Section \ref{sec:mdef} we compare the splashback radius and mass to several commonly used mass definitions, as well as the mass within a fixed multiple of the halo scale radius.  In Section \ref{sec:discussion} we discuss several implications of the existence of the splashback radius, as well as its potential observational signatures and possible detections in existing observations. We summarize our results and conclusions in Section \ref{sec:conclusion}. In Appendix \ref{sec:toy} we present a detailed analysis of various contributions to the growth of halo masses.

Throughout the paper, we denote the mean matter density of the universe at the redshift of analysis as $\rho_{\rm m}$ and the critical density as $\rho_{\rm c}$. Mass definitions using a constant overdensity $\Delta$ relative to $\rho_{\rm m}$ or $\rho_{\rm c}$ are denoted as $M_{\Delta \rm m} = M(<R_{\Delta \rm m})$, e.g. $\mtom$, or $M_{\Delta \rm c} = M(<R_{\Delta \rm c})$, e.g. $\mtoc$. $\mvir$ and $\rvir$ denote masses and radii defined using the redshift-dependent ``virial'' contrast $\Delta(z)$, computed using the approximation of \citet{bryan_98_virial}.

 
\section{The Collapse of Density Peaks and Halo Mass}
\label{sec:peakcollapse}

In this section, we consider the most important processes that occur during the collapse of density peaks in hierarchical structure formation models, and discuss how these processes relate to definitions of the halo boundary and mass. We first establish this connection in the context of simplified analytical models, and then show that the predictions of these models manifest themselves in simulated CDM halos.

\subsection{The Halo Mass and Boundary in the Spherical Collapse Model}
\label{sec:spherical}

In order to elucidate the connection between the collapse of a halo and its boundary, we wish to consider a simple, analytical model. The simplest such model describes the collapse of a top-hat perturbation. However, as laid out in the introduction, peaks in a Gaussian density field have density profiles that decrease with radius, significantly different from a constant density spherical top-hat. While a top-hat perturbation collapses at a well-defined moment in time, the radial shells associated with realistic peaks collapse at different times, resulting in a halo formation history that is extended in time. Thus, we expect better guidance from collapse models that describe the extended collapse of matter.

One such model considers the secondary infall of matter onto a pre-existing overdensity in an $\Omega_{\rm m}=1$ universe \citep{fillmore_goldreich84, bertschinger85}. In this spherical collapse model, all mass shells are bound to the pre-existing overdensity. Each shell will initially expand with the Hubble flow, decelerate, eventually {\it turn around} and start contracting. At some point, the shell will cross previously collapsed shells that are now oscillating in the perturbation potential, thereby entering the {\it multi-stream region} of the halo. The matter of the shell will eventually pass through the pericenter of its orbit in the inner region of the perturbation and expand to the apocenter of its first orbit. Each successive shell collapses onto a deeper potential well than the preceding shell, and thus acquires a higher energy and a larger orbit apocenter. 

In this picture, material piles up near the apocenter due to its small radial velocity in this region of the orbit, creating a density enhancement or {\it caustic} which is extremely sharp in the case of spherical symmetry \citep[see, for example, the detailed discussion in][]{mohayayee_shandarin06}. This caustic occurs at radii $r \approx 0.1-0.4$ times the turnaround radius of the material at apocenter \citep[see, e.g., Fig. 4 of][]{vogelsberger_etal11}, depending on the slope of the density profile of the initial perturbation which determines the mass accretion rate of the collapsing halo. Even in the case of ellipsoidal collapse, the caustic region is marked by a sharp jump in the density profile \citep{adhikari_14}. 

We note that the {\it outermost} caustic corresponds to the apocenter of matter on its {\it first} orbit, i.e. the {\it splashback radius} of the newly accreted matter, which we denote as $\rsp$. The splashback radius cleanly separates the multi-stream region of the perturbation at $r<\rsp$ from the infall region at $r>\rsp$, where successive shells of matter have not yet crossed. The spherical collapse model thus motivates $\rsp$ as a natural definition of the halo boundary, and the halo mass as the mass within this radius, $\msp \equiv M(<\rsp)$. By definition, the increase in $\msp$ between two epochs $z_{\rmi}$ and $z_{\rmf}$  is entirely due to the mass shells that have entered the multi-stream region in the interval $\Delta z=z_{\rmi}-z_{\rmf}$. Therefore, $d\msp/dt$ is the true halo mass growth rate in this model. In contrast, if we had chosen a smaller radius $\rdelta < \rsp$ as the halo boundary, the halo mass growth rate would be due to both the accretion of new matter during $\Delta z$ and matter previously accreted at $z>z_{\rmi}$, as we discuss in detail in Appendix \ref{sec:toy}. 

\subsection{The Mass and Boundary of Realistic CDM Halos} 

\begin{figure*} \centering{
\includegraphics[trim = 8mm 3mm 33mm 0mm, clip, scale=0.625]{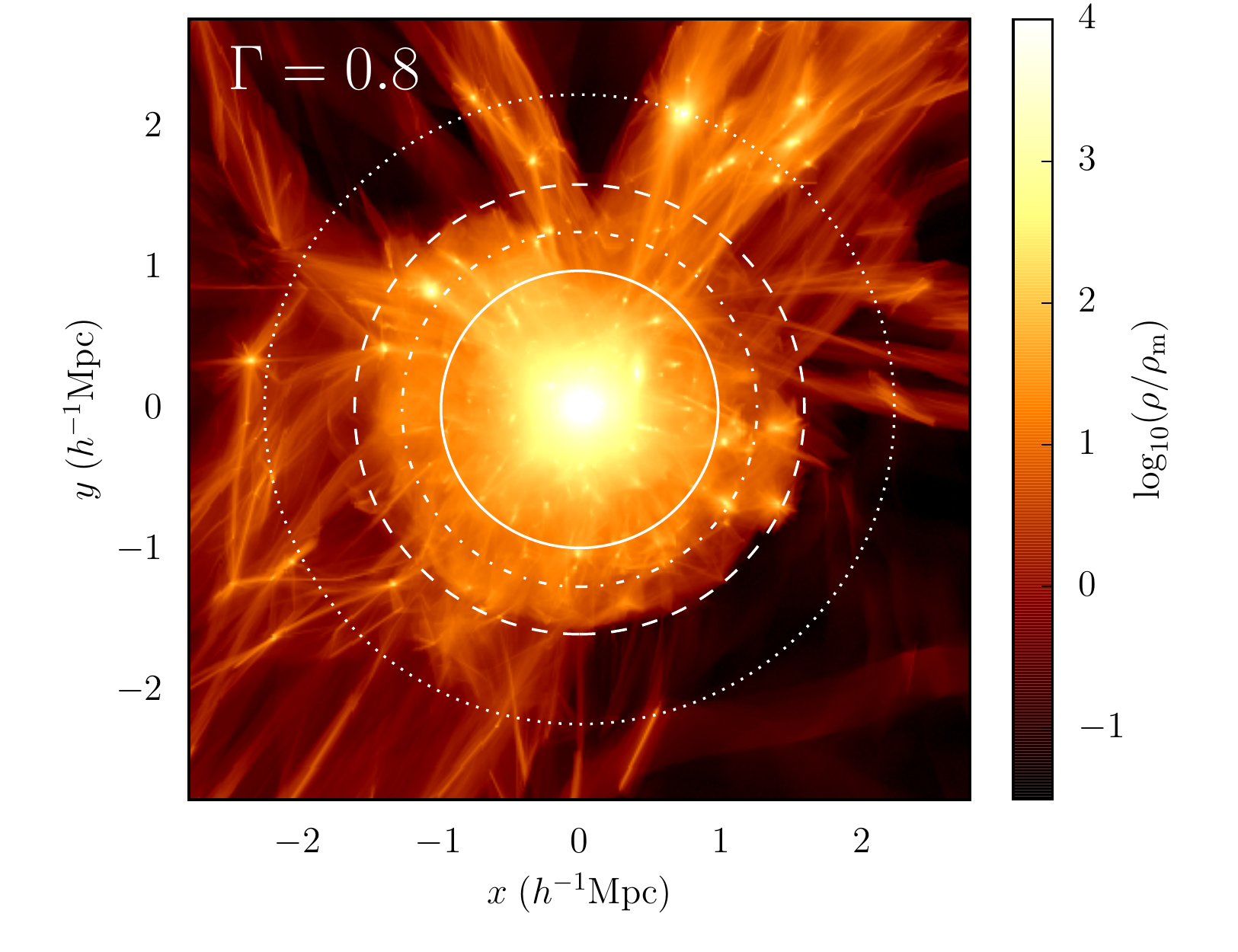}
\includegraphics[trim = 2mm 3mm 0mm 0mm, clip, scale=0.625]{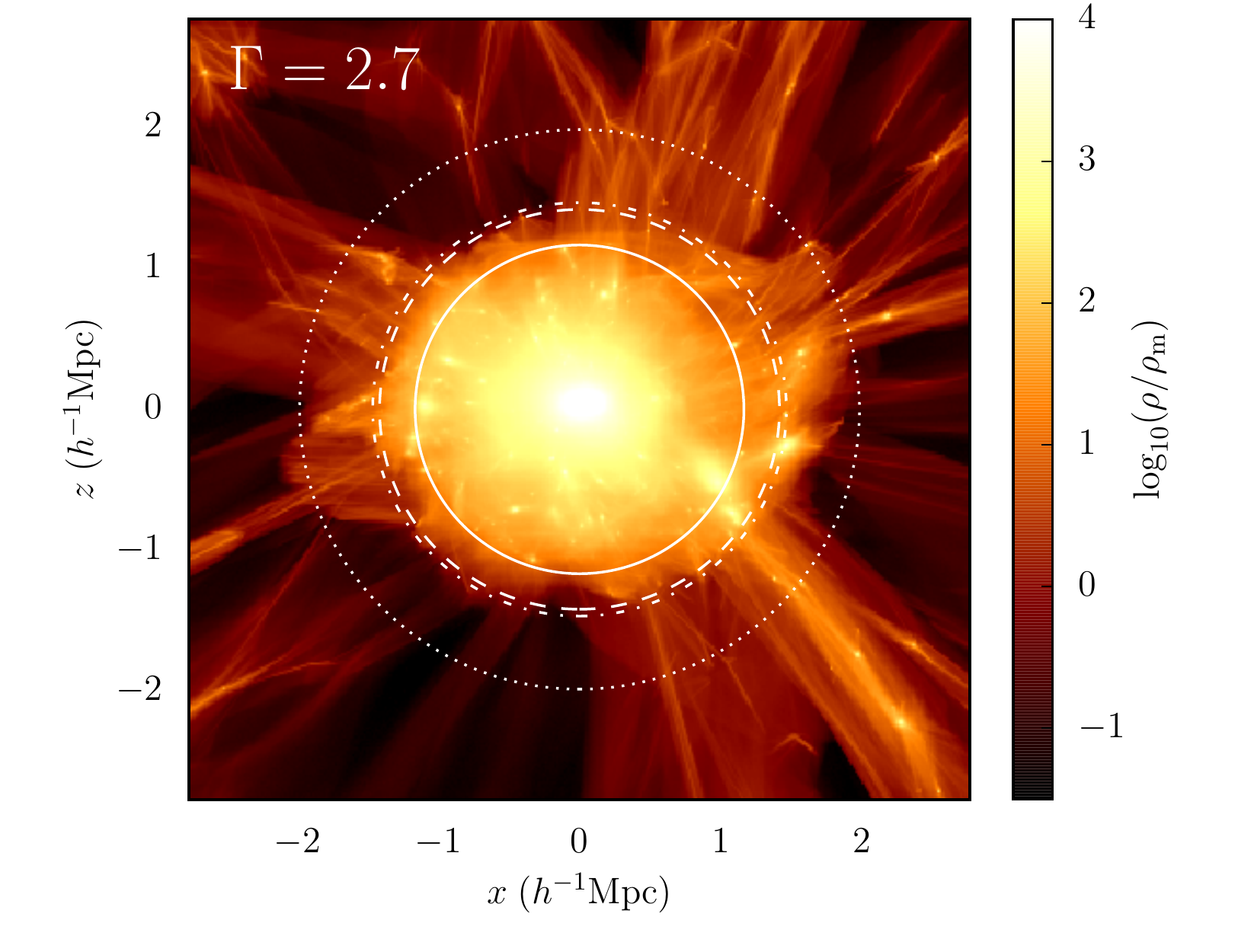}}
\caption{Projected density in a slice of thickness $0.15 \rtom$ through the center of two halos with low (left, $\Gamma = 0.8$) and high (right, $\Gamma = 2.7$) mass accretion rates. The halos have similar masses, $\mvir = 1.1 \times 10^{14}$ and $1.8 \times 10^{14} \msunh$ at $z=0$. The white lines show $\rvir$ (solid), $\rtom$ (dot-dashed), $\rsp$ (dashed) and $\rinfall$ (dotted; see \S \ref{sec:msp} for a detailed description of these radii). $\rsp$ and $\rinfall$ were calculated using the calibrations presented in Section \ref{sec:msp} rather than the density profiles of the individual halos shown. Halos with a low mass accretion rate exhibit a caustic at a radius significantly larger than $\rtom$, whereas fast-accreting halos have $\rsp \lsim \rtom$ (at $z = 0$). The visualizations were created using the algorithm of \citet{kaehler_12}.}
\label{fig:viz}
\end{figure*}

\begin{figure*} \centering{
\includegraphics[trim = 0mm 4mm 0mm 6mm, clip, scale=0.7]{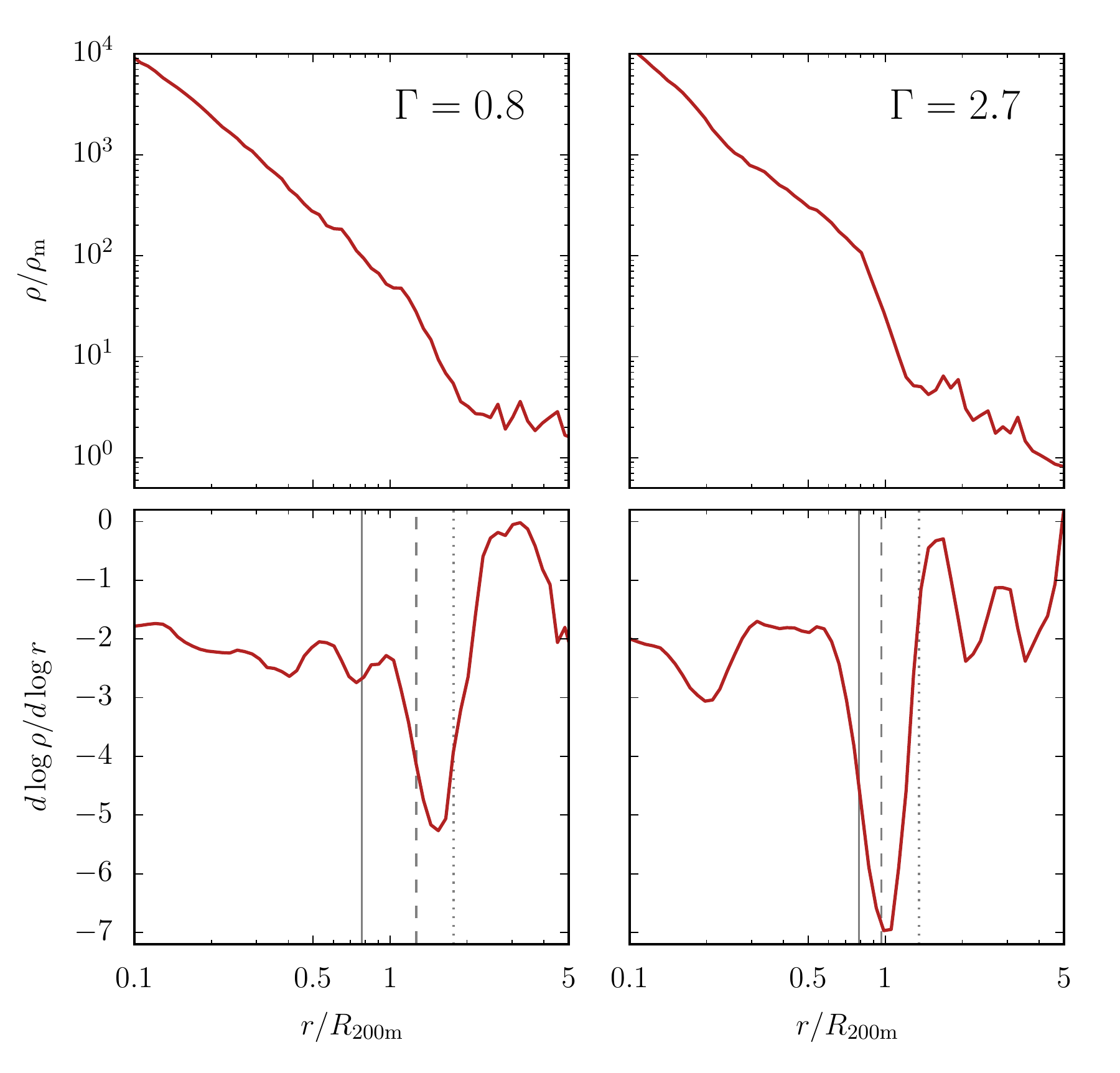}}
\caption{Spherically averaged density profiles (top panels) and their logarithmic slope (bottom panels) of the two halos shown in Figure \ref{fig:viz}. The slopes were computed using a profile smoothed with the fourth-order \citet{savitzky_64} filter over the $15$ nearest bins. The steepening around $\rsp$ is very pronounced in both profiles, but the profile of the faster accreting halo reaches a steeper slope and at a smaller radius. The vertical lines in the bottom panels mark the same radii shown in Figure \ref{fig:viz} using the same line types, i.e. $\rvir$, $\rsp$, and $\rinfall$ (defined as the radius where the mean radial velocity profile of $\bar{v}_r$ reaches minimum) from left to right. For the slower accreting halo (left), the estimate of Equation \ref{eq:rsp_gamma} slightly underestimates the true $\rsp$. This disagreement is not surprising since the $\rsp$ of individual halos are expected to scatter around the median relation.}
\label{fig:profiles}
\end{figure*}

Although the collapse of realistic CDM halos is considerably more complicated than the collapse of a single peak in the secondary infall model, we can still use this model to guide our choices of the halo boundary and mass definitions. As shown by \citet[][see also \citealt{adhikari_14, lau_etal14}]{diemer_14_pro}, halos that accrete mass at a sufficiently high rate do exhibit a sharp steepening of the density profile in the outer regions which is due to the caustic formed by recently accreted matter. The radius at which the profile achieves its steepest slope depends on the halo accretion rate and varies from $\rsp \approx 0.8 \rtom$ for fast-accreting halos to $\rsp \approx 1.5 \rtom$ for slowly accreting halos \citep{diemer_14_pro}. \citet{adhikari_14} have confirmed this result and showed that the location of the steepest slope can be reproduced using the radius of the outermost caustic in the simple model of spherical collapse discussed in Section \ref{sec:spherical}. In particular, they demonstrated that $\rsp$ depends on both the mass growth rate and cosmological parameters such as $\Omega_{\rm m}(z)$.

Figures \ref{fig:viz} and \ref{fig:profiles} illustrate the correspondence between density profiles and $\rsp$ using the example of two individual, cluster-sized halos with similar masses but very different mass accretion rates, representative of the slow and fast accreting sub-populations. We operationally define the mass accretion rate the same way as in \citet{diemer_14_pro},
\begin{equation}
\Gamma \equiv \Delta \log(M_{\rm vir}) / \Delta \log(a) \,.
\label{eq:Gamma}
\end{equation}
Figure \ref{fig:viz} shows the density distribution in a slice through the halo center, while Figure \ref{fig:profiles} shows the spherically averaged density profiles and their logarithmic slope. Both figures contrast $\rsp$ (dashed lines) with $\rtom$ (dot-dashed lines) and the ``virial'' radius $\rvir$ (solid lines). The $\rsp$ radii shown in the figures were predicted using the median relation given by Equation (\ref{eq:rsp_gamma}) below and the $\Gamma$ of the specific halos as determined from the halo catalogs. Figure \ref{fig:viz} clearly demonstrates that the density fields exhibit a sharp jump at $R \approx \rsp$, and that this radius occurs at a smaller multiple of $\rtom$ for the faster accreting halo. We confirm these impressions by considering the spherically averaged density profiles of the same halos in Figure \ref{fig:profiles} which highlights how steep the density profile can get around $\rsp$ (a logarithmic slope of $-7$).

The correlation between $\rsp/\rtom$ and the accretion rate mimics the correlation between the ratio of the last caustic and turnaround radii and the slope of the initial perturbation profile in the spherical collapse model \citep[see][and Section \ref{sec:spherical}]{vogelsberger_etal11}. By further analogy with the spherical collapse model, the splashback radius of CDM halos should include approximately all of the mass ever accreted by a halo. Thus, changes in $\msp$ should always correspond to the {\it current} accretion of new mass, implying that $\msp$ is largely unaffected by pseudo-evolution. In Sections \ref{sec:msp} and \ref{sec:disc:practice} we discuss how $\rsp$ can be measured in cosmological simulations and observations, and how $\msp$ and its evolution relate to conventional spherical overdensity masses.

\section{Halo Radius and Mass evolution}
\label{sec:mdef}

In the previous section we argued that the most natural and physical definition of the mass associated with a density peak is the mass enclosed within the radius of the outermost caustic, $\rsp$, which we now consider in detail. Following the discussion of $\rsp$ and $\msp$ and their calibrations using cosmological simulations, we also consider a halo boundary defined as a constant multiple of the scale radius, $R = 4\rs$, and the corresponding mass, $\mfrs$, which characterizes the mass in the inner regions of halos. We then compare the evolution of halo mass and radius in these definitions to the commonly used ``virial'' mass definitions. 

\subsection{The Splashback Mass, $\msp$}
\label{sec:msp}

\begin{figure} \centering{
\includegraphics[trim = 6mm 23mm 0mm 0mm, clip, scale=0.72]{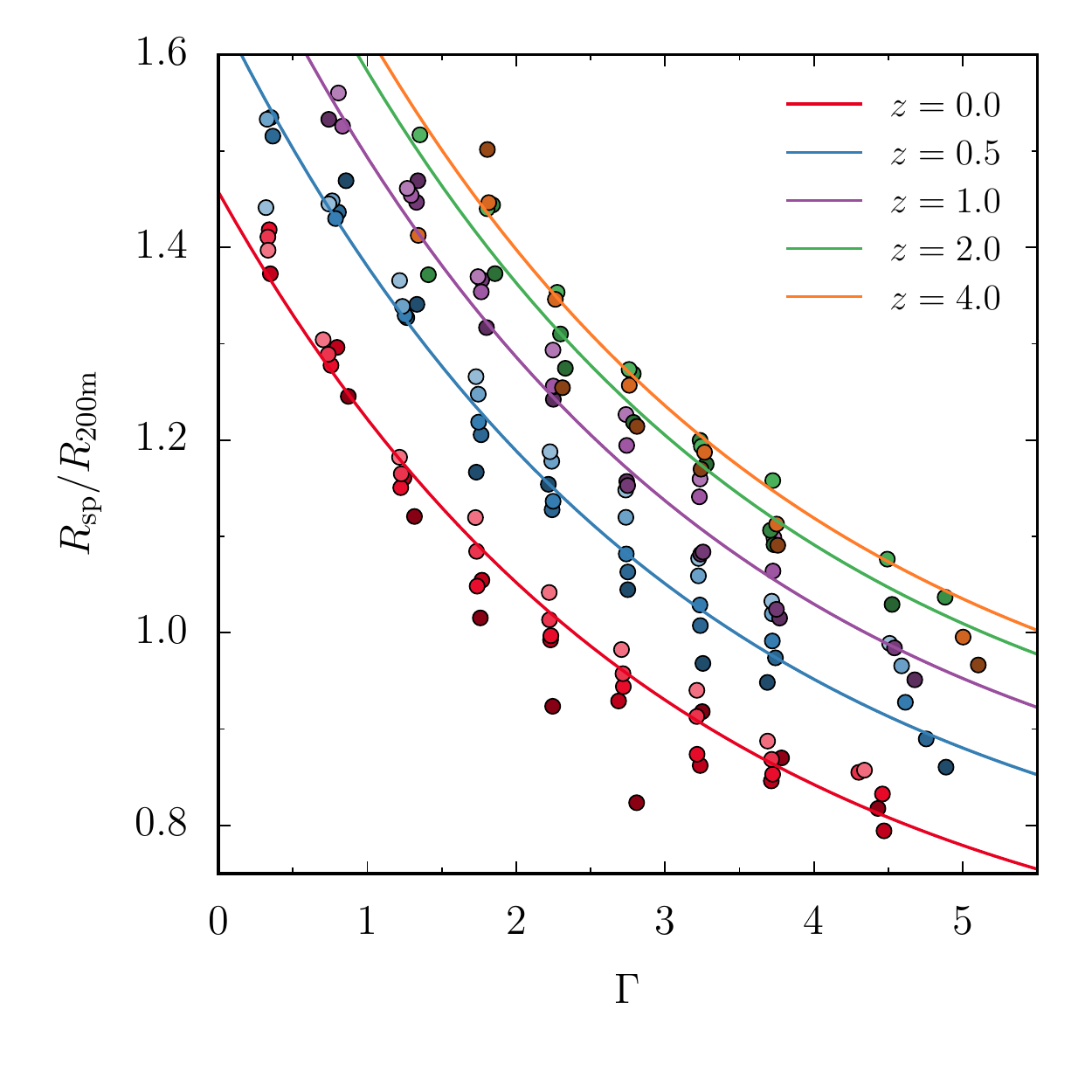}
\includegraphics[trim = 6mm 9mm 0mm 3mm, clip, scale=0.72]{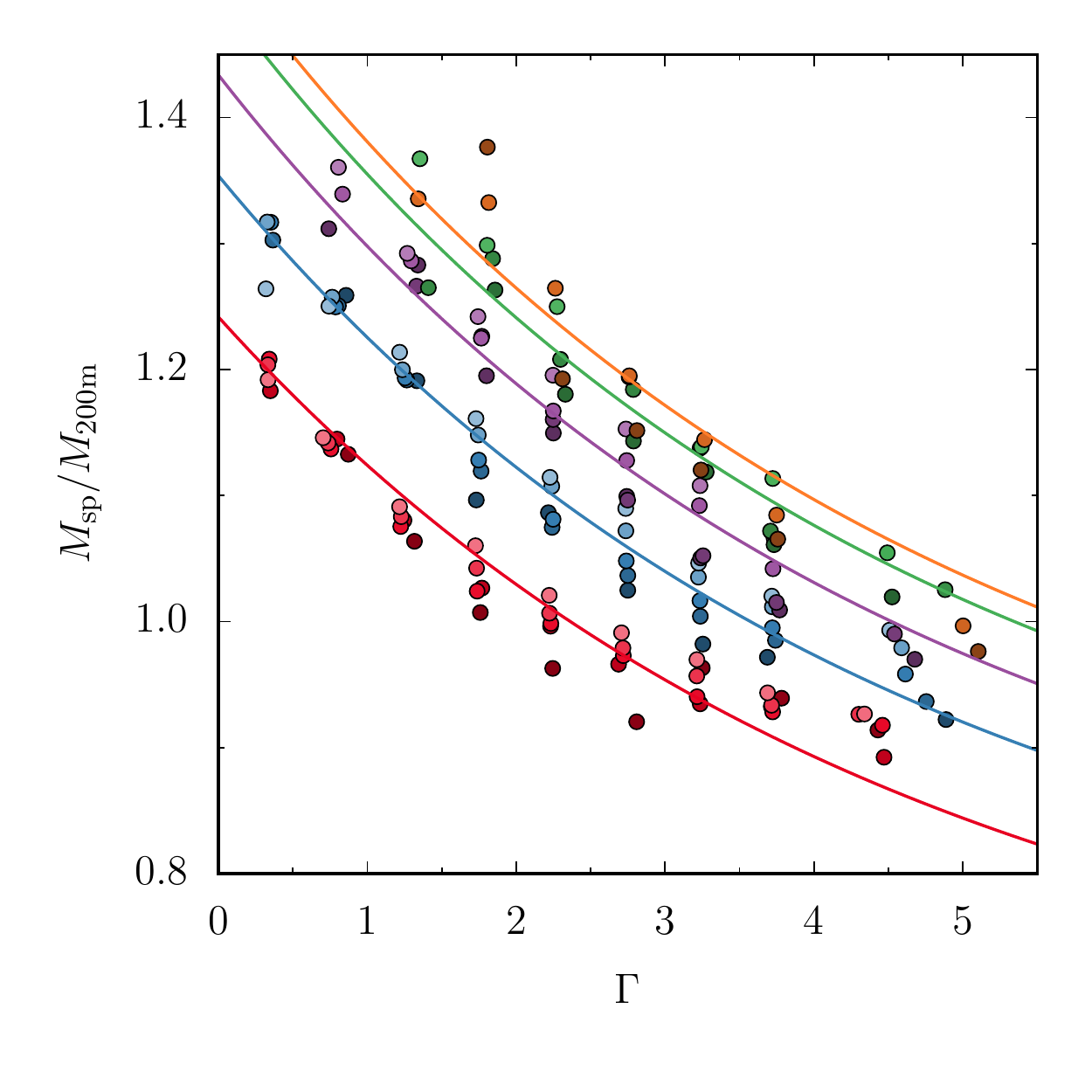}}
\caption{Splashback radius, $\rsp$ (top panel), and the mass within this radius, $\msp$ (bottom panel), as a function of the halo mass accretion rate, $\Gamma$. Darker points correspond to halo samples with higher peak height, $\nu$. The halos were binned in $\nu$-bins of width $0.5$, starting at $\nu = 1$. Samples with $\nu < 1$ were omitted as $\rsp$ is hard to measure for their profiles (see the discussion in Section \ref{sec:msp}). The figure demonstrates that $\rsp / \rtom$ and $\msp / \mtom$ depend on $\Gamma$ and $z$, but do not show a strong dependence on $\nu$ at fixed accretion rate and redshift. For halos in our $\Lambda$CDM cosmology these dependencies can be approximated by Equations (\ref{eq:rsp_gamma}) and (\ref{eq:msp_gamma}), shown with a solid line for each redshift.}
\label{fig:gamma_rsp}
\end{figure}

\begin{figure} \centering{
\includegraphics[trim = 2mm 3mm 0mm 1mm, clip, scale=0.7]{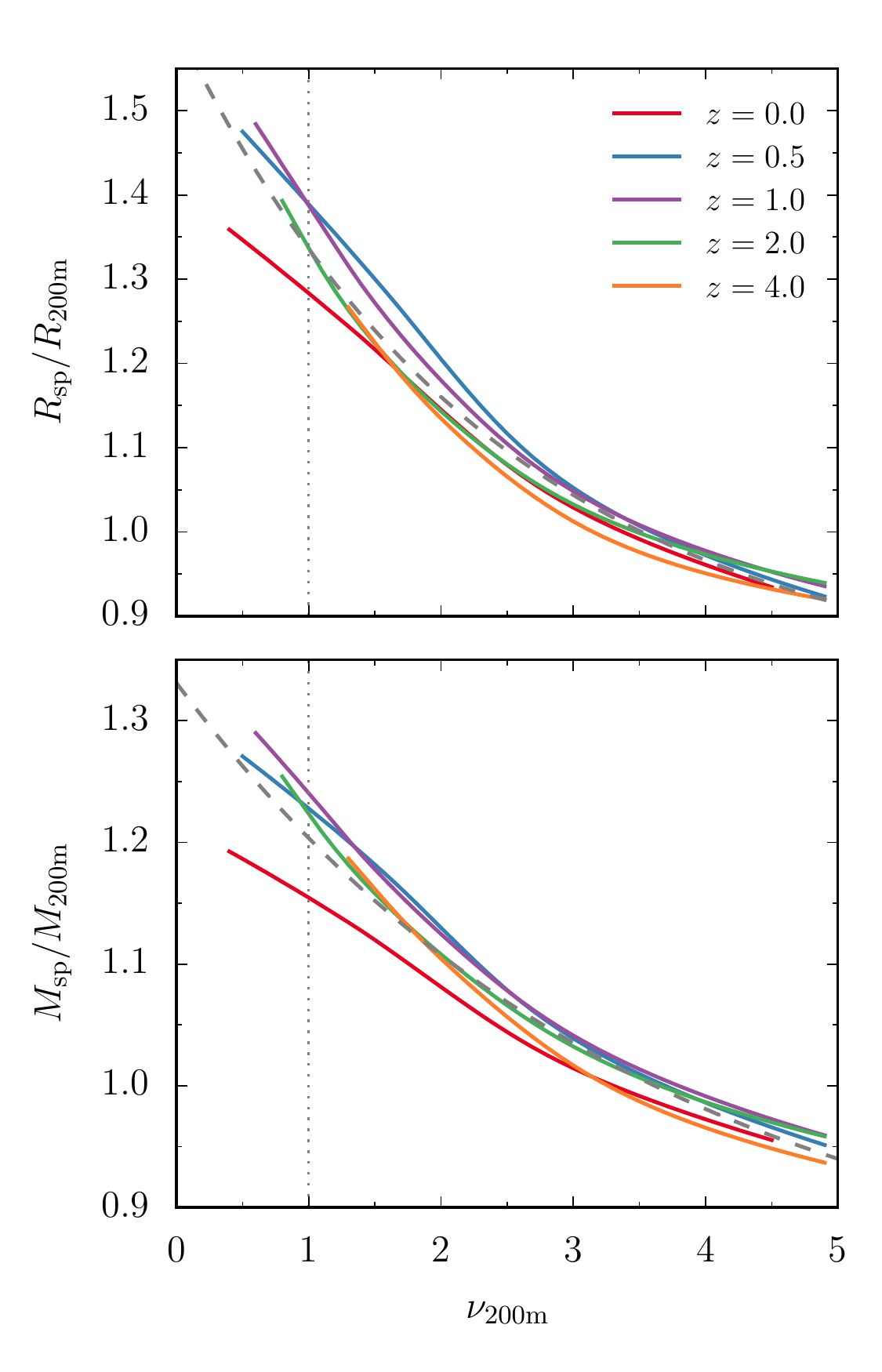}}
\caption{Splashback radius, $\rsp$ (top panel), and the corresponding enclosed mass, $\msp$ (bottom panel), relative to $\rtom$ and $\mtom$ as a function of peak height, $\nu_{\rm 200m}\equiv \delta_c/\sigma(\mtom)/D(z)$. The dotted vertical lines indicate $\nu = 1$, the peak height below which the relation was not directly calibrated (see Figure \ref{fig:gamma_rsp}). The dependence of $\rsp / \rtom$ arises because halos with higher $\nu$ have, on average, a higher mass accretion rate, and thus a smaller $\rsp / \rtom$ (Figure \ref{fig:gamma_rsp}). The dashed lines show the relations specified in Equations (\ref{eq:rsp_nu}) and (\ref{eq:msp_nu}). The figure shows that for the rare, massive halos that accrete mass at a fast rate the splashback radius is close to $\rtom$, while for low-$\nu$ halos it extends to a considerably larger radius of $\rsp\sim 1.5\rtom$.}
\label{fig:nu_rsp}
\end{figure}

As shown by \citet{diemer_14_pro}, $\rsp$ can be measured from the halo density profile as the radius where the density profile steepens sharply beyond what is expected from the NFW and Einasto predictions. We use the same simulation suite and fit the median density profiles of halo samples with a range of different masses, accretion rates, and redshifts using the fitting function in Equation (4) of \citet{diemer_14_pro}, which we reproduce here for completeness,
\begin{eqnarray}
\rho(r) &=& f_{\rm trans}\, \rho_{\rm Einasto} + \rho_{\rm outer} \nonumber \\
\rho_{\rm Einasto} &=& \rho_{\rm s} \exp\left( -\frac{2}{\alpha} \left[ \left(\frac{r}{r_\rms}\right)^{\alpha} - 1\right] \right) \nonumber \\
f_{\rm trans} &=& \left[ 1 + \left(\frac{r}{r_\rmt}\right)^\beta \right]^{-\gamma/\beta} \nonumber \\
\rho_{\rm outer} &=& \rho_\rmm \left[ b_\rme \left( \frac{r}{5\rtom} \right)^{-s_\rme} + 1 \right] \,.
\end{eqnarray}
For consistency, we fix some of the parameters in the fitting function in all fits (regardless of whether the samples were selected by mass or accretion rate), namely $\beta = 6$, $\gamma = 4$. Similar to \citet{diemer_14_pro}, we also fix $\alpha$ according to the relation with $\nu$ as calibrated by \citet{Gao:2008},
\begin{equation}
\alpha(\nu) =  0.155 + 0.0095 \nu^2\,.
\end{equation}
The other parameters, namely $\rs$, $r_{\rm t}$, $b_{\rm e}$, and $s_{\rm e}$, are determined from a least-squares fit. 

\citet{diemer_14_pro} showed that, at $z = 0$, the turnover radius, $r_{\rm t}$, at which the density profiles steepens, depends on the mass accretion rate $\Gamma$. In Figure \ref{fig:gamma_rsp}, we extend this analysis to higher redshifts and use $\rsp$, defined as the radius where median profile of halos reaches the steepest slope, instead of $r_{\rm t}$. The redshift intervals over which $\Gamma$ are measured are the same as the redshifts listed in Figure \ref{fig:gamma_rsp}, i.e. for $z = 0$, $\Gamma$ is measured between $z = 0.5$ and $z = 0$, for $z = 0.5$ between $z = 1$ and $z = 0.5$, and for $z = 4$ between $z = 6$ and $z = 4$. The choice of the redshift intervals defining $\Gamma$ is somewhat arbitrary, but corresponds reasonably closely to the expected crossing time through the full extent of the halo, $2R$ (for example, at $z = 0$, $\Delta z$ corresponds to about 5 Gyr whereas the crossing time is about 4 Gyr).

We bin halos both by $\nu$ and by $\Gamma$, and only use halo samples with $\nu > 1$ ($M > 3 \times 10^{12} \msunh$ at $z = 0$, $M > 10^{11} \msunh$ at $z = 1$) for this analysis, as the density jump associated with the splashback radius is difficult to measure robustly from spherically averaged profiles in halos with low $\nu$ and low $\Gamma$. This issue is apparent in Figure 10 in \citealt{diemer_14_pro}: for profiles with low $\nu$ and low $\Gamma$, the 2-halo term begins to dominate at radii smaller than $\rsp$, thus concealing the steepening in the density profile. This does not mean that low-mass halos do not exhibit a steepening in their density profile; however, they are strongly influenced by their environment, making it difficult to discern the location of $\rsp$.

Figure \ref{fig:gamma_rsp} demonstrates that, at fixed $\Gamma$, $\rsp/\rtom$ does not depend on $\nu$, but does depend on $z$. In particular, \citet{adhikari_14} showed that the overdensity associated with the splashback radius depends on $\Omega_{\rm m}(z)$, where $\Omega_{\rm m}(z) \equiv \rho_{\rm m}(z)/\rho_{\rm crit}(z)$. Thus, we parameterize the dependence of $\rsp/\rtom$ on $\Gamma$ and $z$ with the fitting function 
\begin{equation}
\frac{\rsp}{\rtom} = 0.54 \, \left[1 + 0.53 \Omega_{\rm m}(z) \right] \, \left(1 + 1.36 e^{-\Gamma / 3.04}\right) \,,
\label{eq:rsp_gamma}
\end{equation}
shown with solid lines in Figure \ref{fig:gamma_rsp}. Given this function, we could now compute the median $\msp$ from $\rsp$ by assuming a particular form of the density profile. However, we get a more accurate fit by directly calibrating the median ratio of $\msp$ and $\mtom$ using the simulated density profiles,
\begin{equation}
\frac{\msp}{\mtom} = 0.59 \, \left[1 + 0.35 \Omega_{\rm m}(z) \right] \, \left(1 + 0.92 e^{-\Gamma / 4.54}\right) \,.
\label{eq:msp_gamma}
\end{equation}
These formulae were calibrated using a flat $\Lambda$CDM cosmology where $\Omega_{\rm m}=1-\Omega_{\Lambda}=0.27$, $h=100/H_0=0.7$, $\sigma_8=0.82$, $n_s=0.95$. These calibrations can be used to compute the dependence of the overdensity, $\Delta$, of halos on $\Gamma$ and $\Omega_\rmm(z)$. Since we do not find a large dependence of $\rsp/\rtom$ on $\nu$, we will extrapolate the relations calibrated above even for $\nu<1$ halos in the subsequent sections while discussing our results. We also note that \citet{adhikari_14} have presented a calibration for $\Delta$ as a function of the instantaneous mass accretion rate, $s$ and $\Omega_\rmm(z)$. Once the differences between $s$ and $\Gamma$ are accounted for, our calibration is largely consistent with theirs.

In observations, however, the accretion rate or the exact density profile of a halo are not readily available. Thus, we also quantify the dependence of $\rsp$ and $\msp$ on the conventionally defined, observable $\mdelta$, or rather peak height, $\nu \equiv \delta_c / \sigma(\mdelta) / D(z)$, in Figure \ref{fig:nu_rsp}.\footnote{The peak height is defined using the $\mtom$ mass, i.e. $\nu = \nu_{\rm 200m} \equiv \delta_c / \sigma(\mtom)/D(z)$, where $\delta_c$ is the critical threshold for collapse, $\sigma^2$ is the variance of initial density fluctuations when smoothed with a top-hat filter with a size corresponding to the Lagrangian radius of mass $\mtom$, and $D(z)$ is the growth factor. For the color scale of Figure \ref{fig:gamma_rsp}, we use $\nu_{\rm vir}$ for compatibility with \citet{diemer_14_pro}. Note, however, that the difference between $\nu_{\rm vir}$ and $\nu_{\rm 200m}$ is $\leq 5\%$ for all masses and redshifts. See \citet{diemer_14_pro} for the exact definition of $\sigma(M)$.} This dependence arises because halos of higher peak height exhibit, on average, higher accretion rates \citep[see e.g., Figure 8 in][]{diemer_14_pro}. In order to translate Equations (\ref{eq:rsp_gamma}) and (\ref{eq:msp_gamma}) into functions of $\nu$ rather than $\Gamma$, we use the model of \citet{zhao_09_mah} to calculate halo mass growth histories. For each redshift along an accretion history we compute the accretion rate $\Gamma$ across the same redshift intervals as in \citet{diemer_14_pro} and calculate $\rsp$ and $\msp$ using Equations (\ref{eq:rsp_gamma}) and (\ref{eq:msp_gamma}). Figure \ref{fig:nu_rsp} shows the results as a function of peak height. The relations are more or less independent of redshift, and well fitted by the approximations
\begin{equation}
\frac{\rsp}{\rtom} = 0.81 \, \left(1 + 0.97 e^{-\nu / 2.44}\right)
\label{eq:rsp_nu}
\end{equation}
and
\begin{equation}
\frac{\msp}{\mtom} = 0.82 \, \left(1 + 0.63 e^{-\nu / 3.52}\right) \,,
\label{eq:msp_nu}
\end{equation}
shown with dashed lines in Figure \ref{fig:nu_rsp}. Thus, in the concordance cosmological model, the dependence of $\rsp/\rtom$, $\msp/\mtom$ or the overdensity of halos on $\Omega_\rmm(z)$ is approximately cancelled by the dependence of $\Gamma(\nu)$ on redshift based on the median mass accretion histories of halos. For rare, massive halos that accrete mass at a fast rate (on average), the splashback radius is close to $\rtom$, while for low-$\nu$ halos it extends to a considerably larger radius of $\rsp\sim 1.5\rtom$.

The $\rsp$ and $\msp$ calibrations presented in Equations (\ref{eq:rsp_nu}) and (\ref{eq:msp_nu}) were obtained using the median profiles of halos of a given $\Gamma$ or $\nu$. In observations, however, stacking would result in an average of the density profile, not the median $\rsp$ and $\msp$. We have checked that the $\nu$--$\rsp$ and $\nu$--$\msp$ relations obtained from averaged profiles are almost identical to Equations (\ref{eq:rsp_nu}) and (\ref{eq:msp_nu}).

Given that real CDM halos are not spherical, the density jump associated with the splashback radius occurs at different radii in different directions from the halo center. The corresponding feature in the spherically averaged density profile is thus not a sharp jump but rather a steepening of the profile that spans a range of radii. The finite radial extent of the steepening creates a certain ambiguity in the choice of the splashback radius definition. For instance, an alternative definition of the splashback radius could be the radius where the average radial velocity in a shell is most negative, $\rinfall$. This radius is more likely to include most of the accreted mass, although the majority of the mass between $\rsp$ and $\rinfall$ is infalling for the first time. We find that $\rinfall \approx 1.4 \rsp$ and $\minfall \approx 1.2 \msp$ at all redshifts and halo masses. Thus, the fitting formulae in Equations (\ref{eq:rsp_nu}) and (\ref{eq:msp_nu}) can easily be modified to return $\rinfall$ and $\minfall$. 

Figure \ref{fig:viz} above  shows both $\rsp$ and $\rinfall \approx 1.4 \rsp$ (dotted line). In contrast to $\rsp$, $\rinfall$ clearly extends into the filamentary regions and is not associated with the collapsed halo matter. Figure \ref{fig:profiles} confirms this impression, as $\rinfall$ does not correspond to a particular feature in the density profiles. Thus, the splashback radius definition based on the steepest density profile slope is preferable, and will be used for the remainder of this paper. 

\subsection{The Inner Mass, $\mfrs$}
\label{sec:M4rs}

\begin{figure*} 
\centering{
\includegraphics[trim = 2mm 1mm 2mm 2mm, clip, scale=0.8]{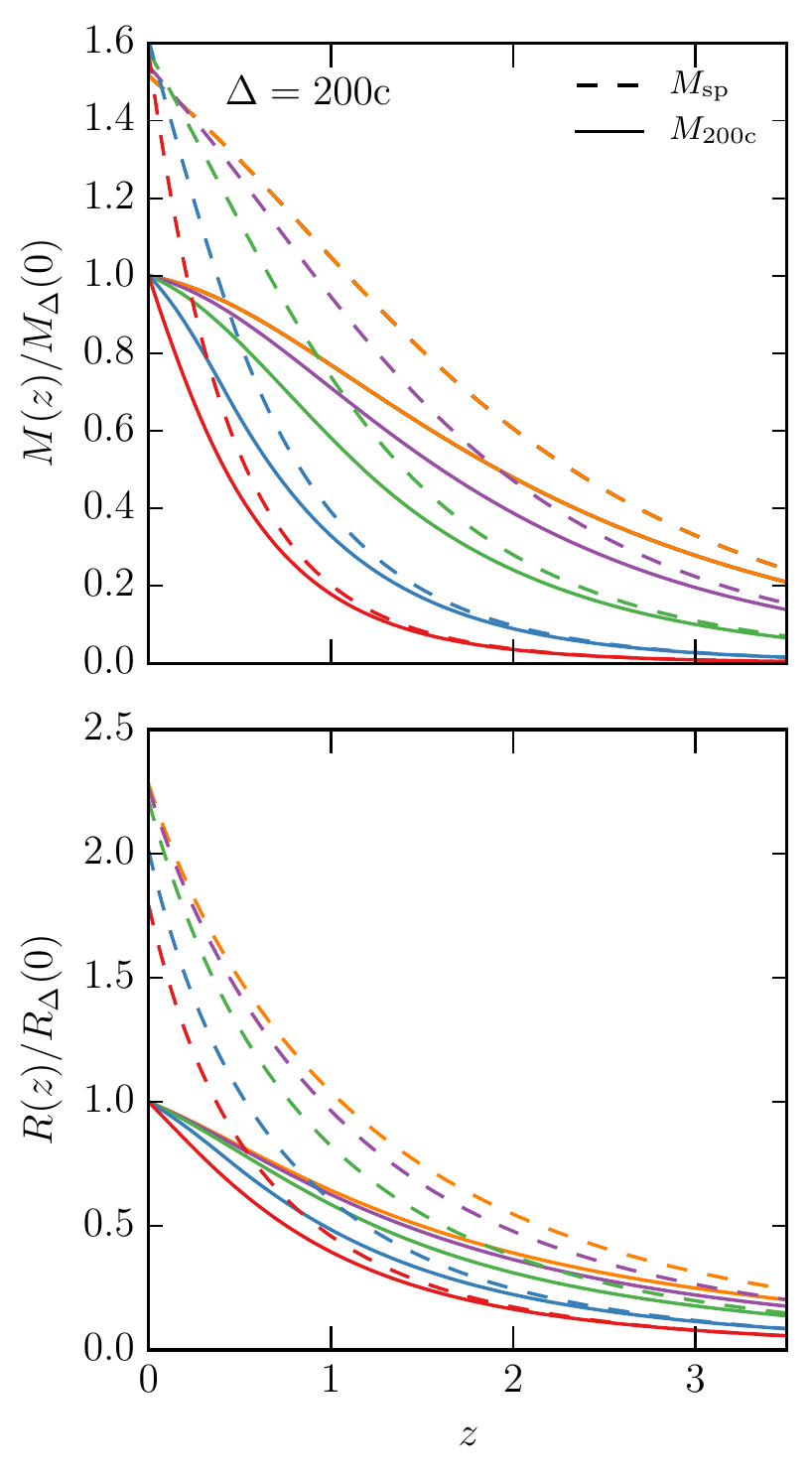}
\includegraphics[trim = 14mm 1mm 2mm 2mm, clip, scale=0.8]{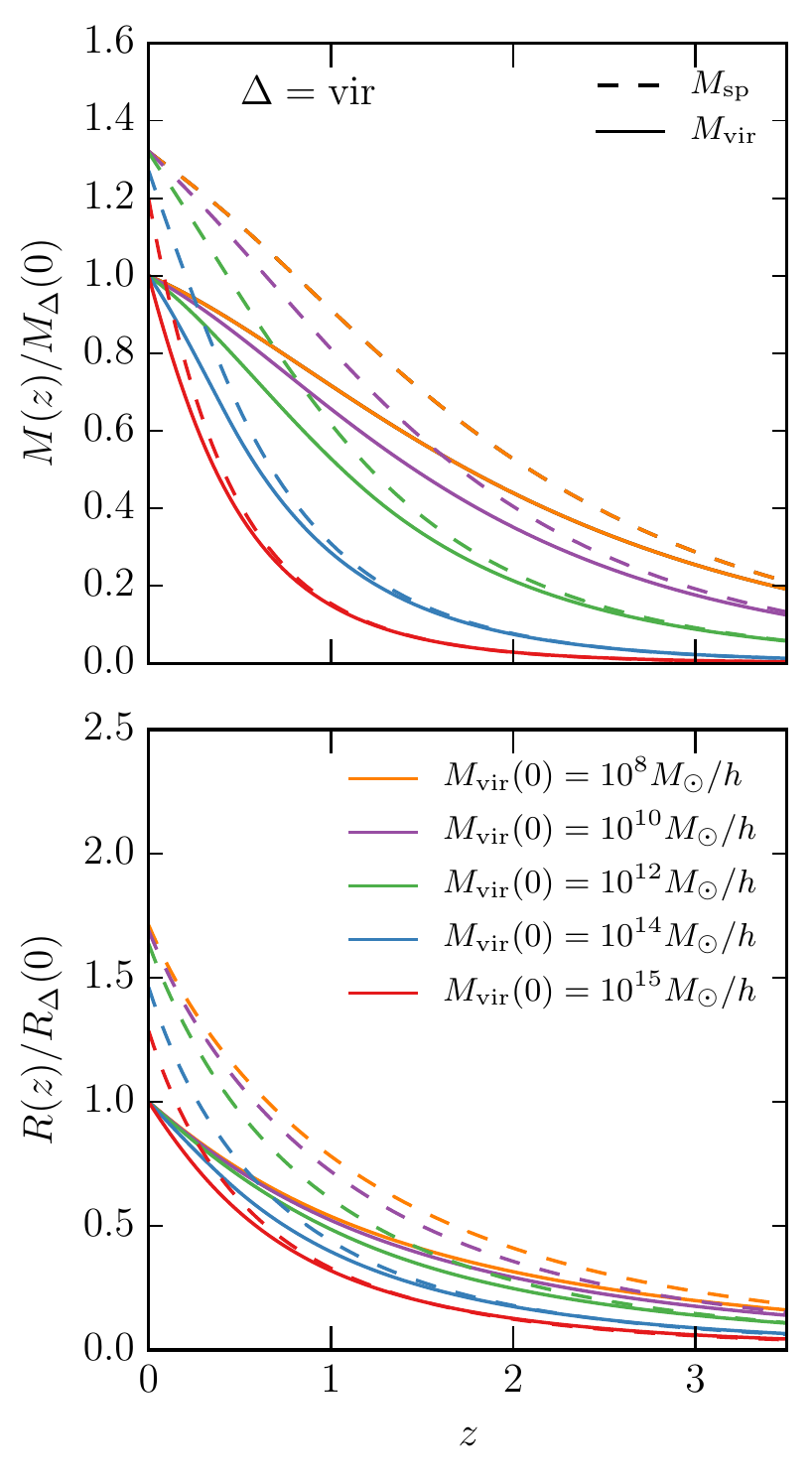}
\includegraphics[trim = 14mm 1mm 2mm 2mm, clip, scale=0.8]{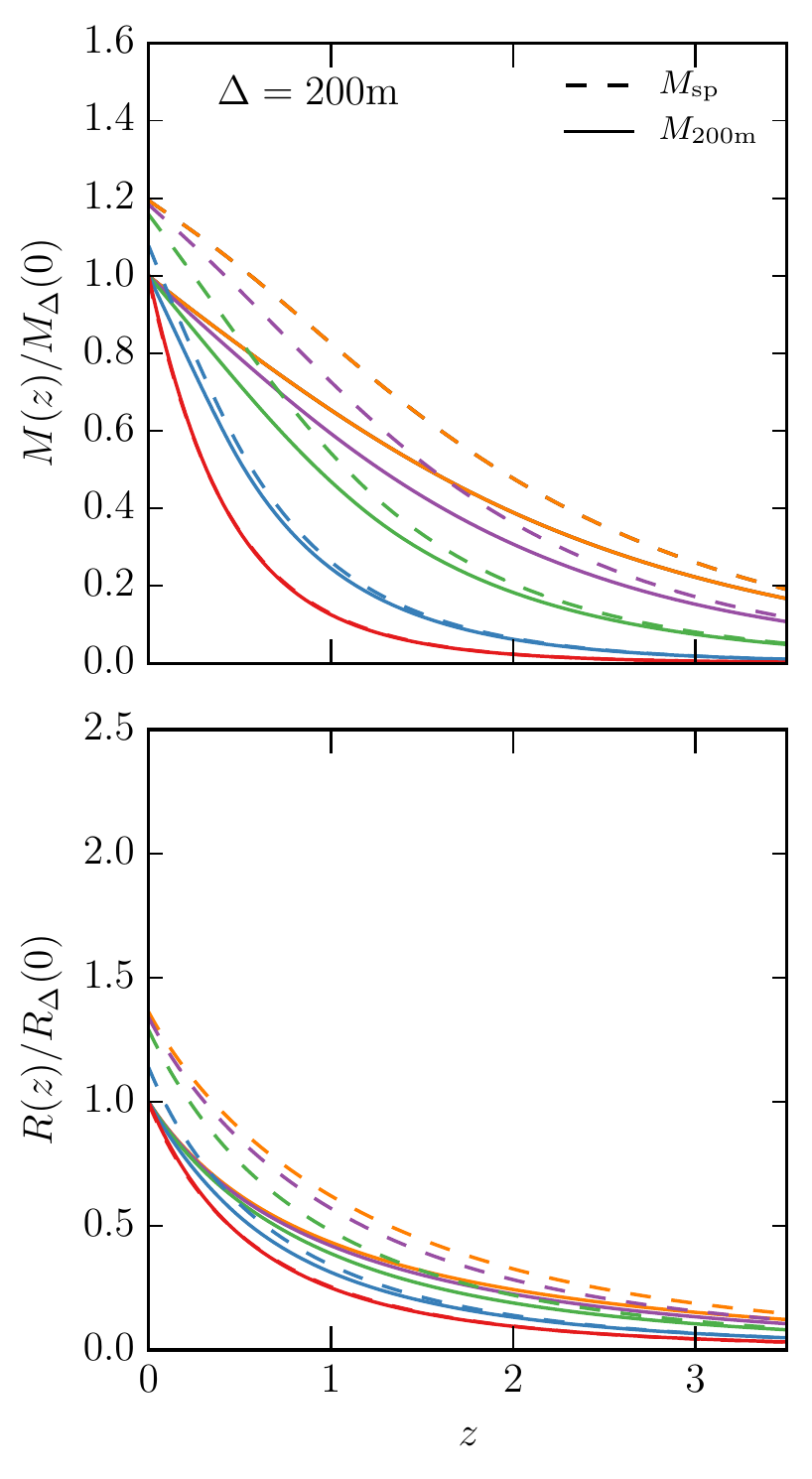}
}
\caption{
Median mass growth histories (top) and halo boundary growth rates (bottom) of halos of different masses $\mdelta$ at $z=0$ (solid lines), as well as the mass within the splashback radius (dashed lines), calculated as described in the text. In the left column we use $\mdelta = \mtoc$, in the middle column $\mdelta = \mvir$, while in the right column $\mdelta = \mtom$. In the fast mass growth regime, $\msp \approx \mvir$, while in the slow mass growth regime $\msp$ grows faster than $\mvir$. The transition between these two regimes happens earlier for smaller halos. The differences between $\msp$ and other mass definitions are small at high redshift, but increase with decreasing redshift. For low mass halos, the high contrast mass definitions such as $\mtoc$ can show differences of up to $50\%$ at $z=0$. For cluster mass halos $\msp \approx \mtom$. The differences in the halo boundaries are much larger than those in the halo masses.
}
\label{fig:mvirmsp}
\end{figure*}

The results presented in Section \ref{sec:msp} demonstrate that the splashback radius of halos is quite large. For some purposes, it may be instructive to consider the mass evolution in the inner regions of halos. In principle, one could characterize the inner regions simply by using $\mdelta$ with a high value of $\Delta$. However, given that any spherical overdensity radius $R_{\Delta} < \rsp$ is subject to pseudo-evolution, we would prefer to define the inner mass using a radius that is not tied to any cosmological reference density. 

\citet[][]{zemp_14} discuss several alternative halo mass and radius definitions that are not subject to pseudo-evolution. In particular, it was argued that the mass within a radius that encloses a fixed physical density independent of redshift, such as $200\rho_{\rm m}(z=0)$, could be used as an alternative measure of mass. Such a definition, however, has a number of drawbacks. First, if the threshold density is chosen to be too low, the corresponding radius will be much larger than the virialized region of halos at high $z$. If the density is chosen to be too high, the enclosed mass will correspond only to the inner region of the halo. Most importantly, the mass and radius defined in this way do not track the physical growth of a halo in its fast accretion regime, where the halo profile does exhibit a well-defined characteristic scale close to $\rtom(z)$ \citep[][see also Figure~\ref{fig:mvirmsp}]{diemer_14_pro}. \citet{zemp_14} has also discussed the possibility of using the scale radius (the radius where the density profile has a logarithmic slope of $-2$) as the halo boundary, and define the mass as $M(<\rs)$.

Here we use a halo radius and mass definition where $R = 4\rs$ and $\mfrs \equiv M(<4 \rs)$.  The multiple of 4 in this definition is motivated by the fact that the concentration, $c_{\Delta}=\rdelta/\rs$, is approximately equal to four as long as the halo is in the fast accretion regime \citep{zhao_03, zhao_09_mah}.\footnote{The concentration in the fast accretion regime has a residual dependence on halo mass and redshift and thus varies between $c_{\rm min} \approx 3$ and $c_{\rm min} \approx 4$  \citep{diemer_14_cnu}. However, for the sake of simplicity, we choose a fixed value of $c_{\rm min} = 4$.} In this regime, where the contribution of pseudo-evolution to the mass growth is relatively small, $\mfrs$ approximately tracks $\mdelta$. Subsequently, as the mass growth and physical evolution of the inner region of the halo profile slow down, the scale radius approaches a constant \citep{bullock_01_profiles} and $\mfrs$ tracks the actual evolution of the inner halo mass due to real profile changes, unaffected by pseudo-evolution. Assuming an NFW density profile, $\mfrs$ is given by
\begin{equation}
\mfrs = M_{\Delta}\frac{\mu(4)}{\mu(c_{\Delta})}\,.
\label{eq:m4rs}
\end{equation}
where $\mu(x)=\ln(1+x)-x/(1+x)$ and the mass and concentration could correspond to any of the commonly used density contrast choices. 

For example, for a Milky-Way sized halo of mass $\mvir = 10^{12} \msunh$ ($\rvir = 207 \kpch$) and a typical concentration at that mass, $\cvir \approx 9$ \citep[e.g.,][]{zhao_09_mah, diemer_14_cnu}, $4\rs \approx 92 \kpch$, and $\mfrs \approx 5.8 \times 10^{11} \msunh$. For comparison, the median $\rsp$ for such a halo is $358 \kpch\approx 511$ kpc and the median $\msp$ is $1.4 \times 10^{12} \msunh$, about $2.4$ times larger than $\mfrs$. We caution that these values are medians, and there is large scatter both in the concentration-mass relation and in $\rsp$ at fixed mass. 

\subsection{The Evolution of Halo Radii and Masses in Different Definitions}
\label{sec:mrcomparison}

We now contrast the redshift evolution of $\rsp$ and $\msp$ with the evolution of the boundary and mass for the definitions $\mfrs$ and $\mdelta$, using some common choices of $\Delta$. Once again, we use the model of \citet{zhao_09_mah} to calculate concentrations and halo mass growth histories. The concentrations are used to convert between the different choices of $\Delta$, while the halo mass growth histories are also used to derive $\Gamma$ (and thus $\rsp$ and $\msp$ using the fitting formulae presented in Section \ref{sec:msp}). 

The upper panels of Figure~\ref{fig:mvirmsp} show the evolution of $\msp$ and the traditional definitions $\mtoc$, $\mvir$ and $\mtom$. At high $z$ (i.e., in the fast mass growth regime), $\msp \approx \mtoc\approx \mvir \approx \mtom$. At low $z$, $\msp$ evolves faster than $\mvir$ for halos of all masses. In particular, the figure demonstrates that {\it even dwarf and Milky Way-sized halos do accrete new mass at low redshifts}. In the lower panels, we compare the evolution of the corresponding radii.  While $\rsp$ can be significantly larger than $\rtom$ (see also Figures \ref{fig:gamma_rsp} and \ref{fig:nu_rsp}), only a relatively small fraction of the total halo mass resides at those radii, reducing $\msp / \mtom$ compared to $\rsp / \rtom$.

\begin{figure} \centering{
\includegraphics[trim = 2mm 1mm 2mm 2mm, clip, scale=0.85]{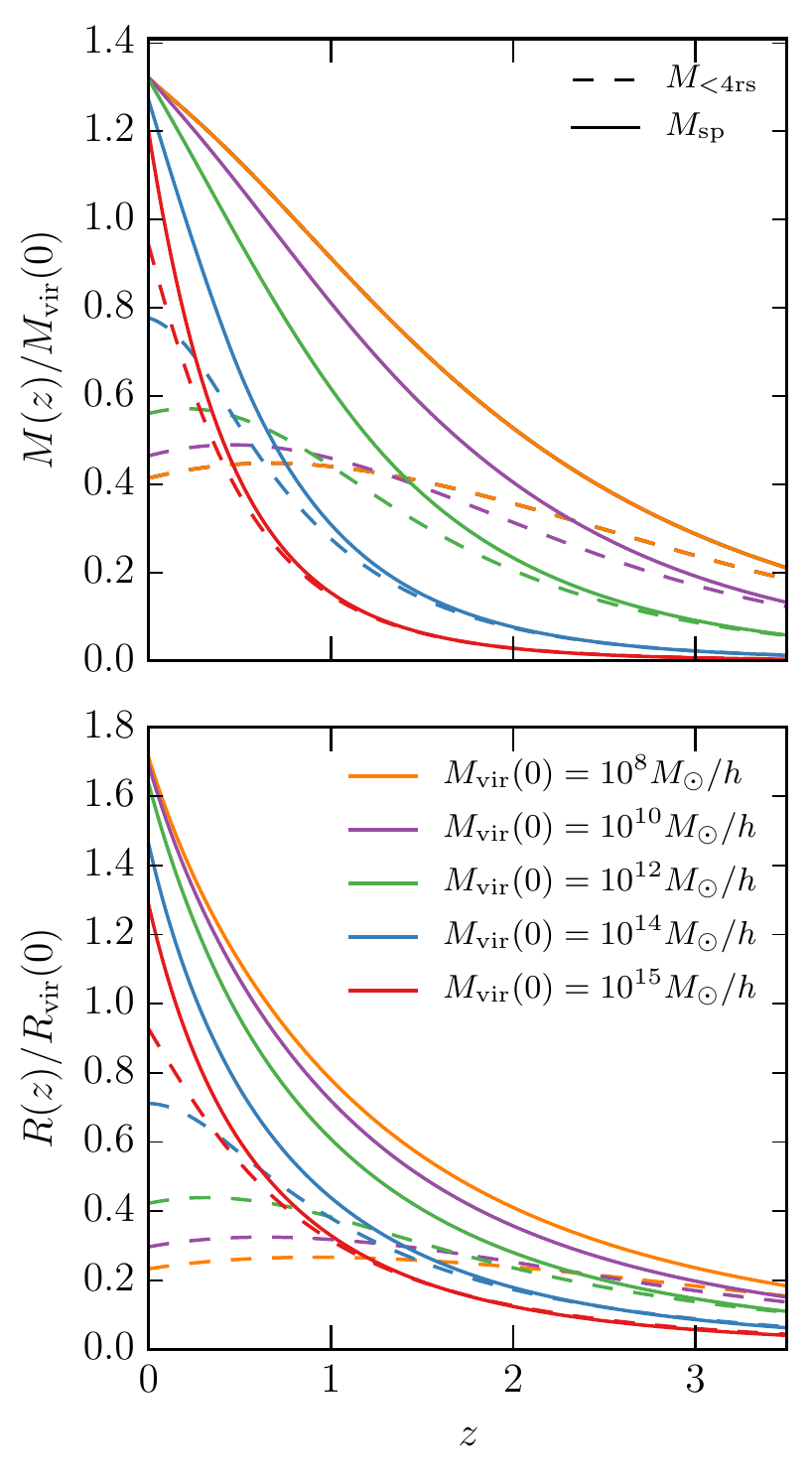}}
\caption{
Median mass growth  (top) and halo boundary growth histories (bottom)
of halos of different virial masses at $z=0$ (solid lines) for the splashback
definition, as well as for the mass within 4 scale radii, $\mfrs$ (dashed
lines), calculated using the model of \citet{zhao_09_mah}. In the fast growth
regime, $\mfrs \approx \mvir \approx \msp$, while in the slow growth regime $\mfrs$ approaches a constant value. The transition between these two regimes happens earlier for smaller halos. The slight decrease in $\mfrs$ at low $z$ appears to be a small artefact of the \citet{zhao_09_mah} model and is not present when we plot a similar evolution using halos from cosmological simulations.}
\label{fig:mvirmform}
\end{figure}

\begin{figure} 
\centering{
\includegraphics{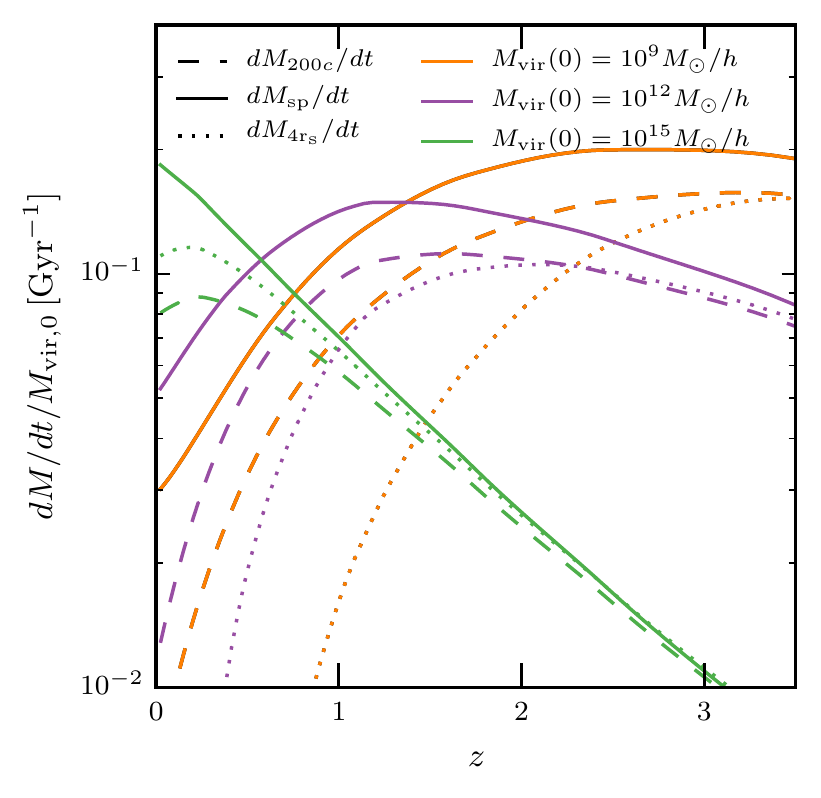}}
\caption{
Comparison of the growth rates of the splashback mass, $\msp$, $\mfrs$, and $\mtoc$. At low
redshifts the difference between the rates based on $\mtoc$ ($\mfrs$) and $\msp$
can be as large as a factor of $2$ ($10$) for low-mass halos.
}
\label{fig:dM200cdt}
\end{figure}

The differences in the growth rates of $\msp$, $\mtoc$, $\mvir$, and $\mtom$ arise because the mass accreted within $\rsp$ at late times is distributed with an approximately isothermal $\rho \propto r^{-2}$ profile and contributes significantly only at large radii, $r\gtrsim \rvir$ (see Section \ref{sec:disc:quiet}). This highlights an important point: at low redshift, {\it the halo mass distribution in the inner regions may be relatively stable and evolve slowly, while the outer regions may evolve fast.} 

It is therefore also interesting to contrast the evolution of $\msp$ with $\mfrs$ which characterizes the mass distribution in the inner regions of halos at low $z$ (see Section \ref{sec:M4rs}). This comparison is shown in Figure \ref{fig:mvirmform}, both for the masses (top panel) and radii (bottom panel). The most massive halos are largely still in the fast accretion regime today, and the evolution of $\msp$ and $\mfrs$ are quite similar. Low-mass halos, on the other hand, are in the slow mass growth regime and their $\msp$ and $\mfrs$ evolve quite differently. For example, for halos of $\mvir(z=0) = 10^8 \msunh$, $\mfrs$ evolves significantly slower than $\msp(z)$ at $z\lesssim 3.5$. At $z\lesssim 1$, $\mfrs$ is approximately constant, while $\msp$ for these halos changes by $\approx 30\%$ between $z=1$ and $z=0$. Finally, we compare the halo mass growth rates for the $\msp$, $\mtoc$ and $\mfrs$ definitions in Figure~\ref{fig:dM200cdt}. The growth rate of $\mfrs$ is approximately an order of magnitude lower than that of $\msp$ at $z \lesssim 1$ for low-mass halos.

In summary, Figures \ref{fig:mvirmsp}-\ref{fig:dM200cdt} show that during the fast accretion regime $\mvir$, $\mtom$, $\msp$, and $\mfrs$ are approximately equivalent. In the slow mass growth regime, however, the mass within the inner radii ($r<r_{4\rs}$) for halos with $M_{\rm vir 0}\lesssim 10^{12} \msunh$ stops growing, while $\msp$ keeps growing even faster than $\mvir$. 

\section{Discussion}
\label{sec:discussion}

In the previous sections, we have argued that a natural definition of the halo boundary is the splashback radius, which corresponds to the apocenter of matter on its first orbit after accretion. Although the distribution of the apocenters of recently accreted material is not spherical as in the idealized models of secondary accretion that motivate the concept, real CDM halos do exhibit a steepening of their radial density profiles over a relatively narrow range of radii. We define the splashback radius as the radius of the steepest slope of the density profile and calibrated this radius and the mass enclosed within it for the concordance cosmological model. Using the splashback radius as a halo boundary has a number of implications.

For instance, the fact that $\rsp / \rtom$ and $\msp / \mtom$ are independent of redshift at a fixed $\nu$ (Figure \ref{fig:nu_rsp}) may have implications for the universality of the halo mass function. The mass function is found to be most universal when masses are defined with respect to the mean density of the universe, such as $\mtom$, and less universal when masses are defined with respect to the critical density. From general considerations, one may expect peak collapse process and the associated halo mass function for $\msp$ to be approximately universal. The universality of $\msp / \mtom$ as a function of $\nu$ would then explain why the halo mass function for the $\mtom$ definition is approximately universal, or at least why it is considerably more universal than for the $\mtoc$ definition.

We have shown that, at late times, the evolution of $\rsp$ and the enclosed mass, $\msp$, is considerably faster than that of $\rtoc$ or $\rvir$ and their corresponding masses. However, $\mfrs$, which characterizes the mass distribution in the inner regions of slow accreting halos, evolves very little at $z\lesssim 1$. This difference implies that a quiescent evolution of the inner regions of halos can co-exist with active growth in the outer regions. Less massive halos at low redshifts, in particular, do accrete mass, even though their inner regions bear scant evidence for such growth. We have also shown that for slow accreting halos at low redshifts the splashback radius can be up to a factor of two larger than $\rtoc$, which is often used to define the halo boundary in galaxy formation studies. This raises the possibility that the ``zone of influence'' of individual halos (where satellite-specific environmental effects can be seen) may be much larger than is usually assumed. In the remainder of this section, we discuss these issues, as well as operational definitions of $\rsp$ and $\mfrs$ in simulations and observations and possible observational detections of the splashback radius. 

\subsection{Quiet Interiors with Active Outskirts}
\label{sec:disc:quiet}

Figures \ref{fig:mvirmsp} and \ref{fig:mvirmform} show that $\msp$ grows significantly all the way to $z=0$ for halos of all masses, while the mass in the inner regions (characterized by $\mfrs$) evolves considerably slower at $z\lesssim 1-2$, and not at all for galaxy-sized halos. The reason for the slow evolution of the inner regions is not that the overall accreted mass is small (galaxy-sized halos approximately double their $\msp$ between $z=1$ and $z=0$; see Figure \ref{fig:mvirmsp}), but that the newly accreted matter has a relatively shallow density profile \citep[see Section 4.2 in][]{lithwick_dalal11}. The shallow profile arises because the radial profile of the potential over most of the halo volume is shallow, meaning that the velocity of the accreted matter does not vary strongly with radius. Hence, the time spent at each radius $r$ is $\delta t(r)\sim r/v\propto r$. The time averaged mass profile of newly accreted mass is thus $M_{\rm acc}(<r)\propto \delta t(r)\propto r/v\propto r^{\alpha}$ with $\alpha\sim 1$, leading to a density profile which is close to isothermal, $\rho\propto r^{-2}$. This profile is shallower than the overall NFW-like profile of CDM halos, which implies that the previously accreted matter dominates in the inner regions of halos while newly accreted mass contributes significantly in the outer regions near the splashback radius. 

This highlights the possibility that systems that appear quiescent in their interior regions may actually still actively grow, particularly when much of the mass growth is due to the accretion of diffuse mass and small halos rather than due to major mergers. The orbit of this material may take it to the interior regions, but it is predominantly deposited in the outskirts. Let us consider some implications for both massive, cluster-sized halos as well as galaxy-sized halos.

\begin{figure} \centering{
\includegraphics{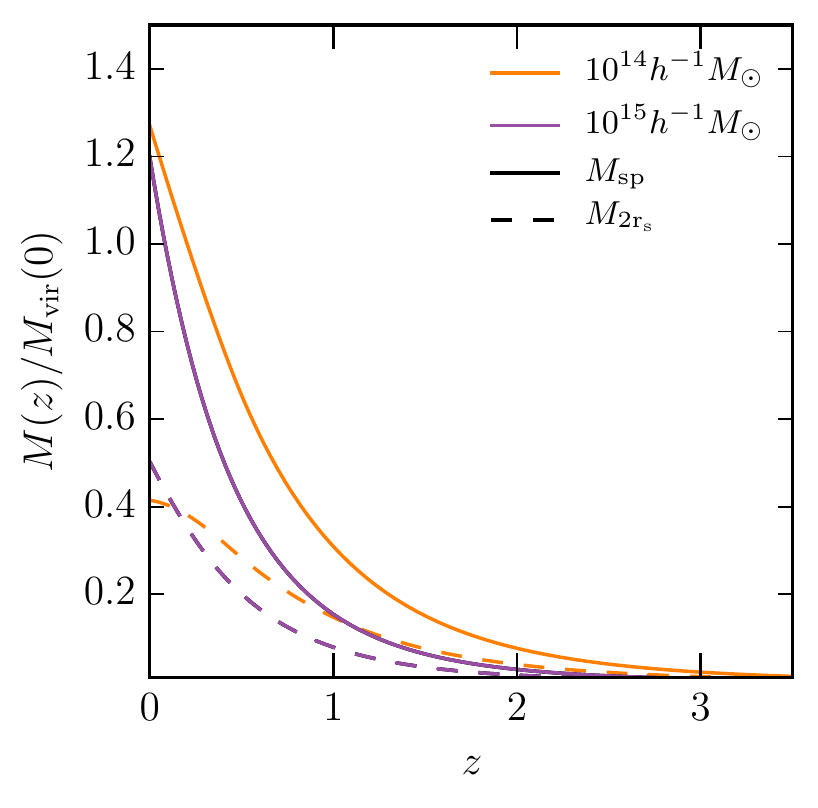}}
\caption{
Median mass growth  (top) and 
of cluster-sized halos for mass within two NFW scale radius, $M_{\rm 2r_s}$ (dashed lines) and mass within  the splashback
radius (solid lines) calculated using the model of for mass accretion history and concentrations \citet{zhao_09_mah}.  The figure shows that 
for $10^{14}\ \msun$ halos $M_{\rm 2r_s}$ grows from $z=1$ to $z=0$ by a factor of two, while $\msp$ grows by a factor of four. For $10^{15}\ \msun$ halos which grow at a much faster rate, the growth of the two masses over the same redshift interval is comparable.}
\label{fig:m2rsmsp}
\end{figure}

For cluster-sized halos, relaxed ``cool-core'' inner regions do not preclude the possibility of active accretion in the outer regions. Abell 133 (further discussed in Section \ref{sec:disc:detections}) may be an example of such a system, as it combines the characteristics of a cool core system (albeit a disturbed one) and a system undergoing rapid mass accretion. We indicate this possibility in Figure \ref{fig:m2rsmsp} which contrasts the average evolution of mass (calculated using the \citealt{zhao_09_mah} model) within two NFW scale radii, $M_{\rm 2r_s}$, and mass within the splashback radius, $\msp$, for cluster-sized halos. We show the mass within $2r_s\approx 2R_{\rm 200c}/c_{\rm 200c}\approx 0.5R_{\rm 200c}\approx R_{\rm 500c}$ because this radius is often used to measure halo masses in cluster studies. The figure shows that for $10^{14}\ \msunh$ halos $M_{\rm 2r_s}$ grows from $z=1$ to $z=0$ by a factor of two, while $\msp$ grows by a factor of four. For $10^{15}\ \msunh$ halos which grow at a much faster rate, the growth of the two masses over the same redshift interval is comparable, which indicates a significant growth of both the inner and outer regions of the halo. We conclude that individual cluster-sized halos can exhibit very different types of mass evolutions: for some halos the difference between the growth of $M_{\rm 2r_s}$ and $\msp$ is more than a factor of two, while for other, fast growing halos the difference may be small.
 
\begin{figure} \centering{
\includegraphics[trim = 1mm 2mm 2mm 1mm, clip, scale=1.0]{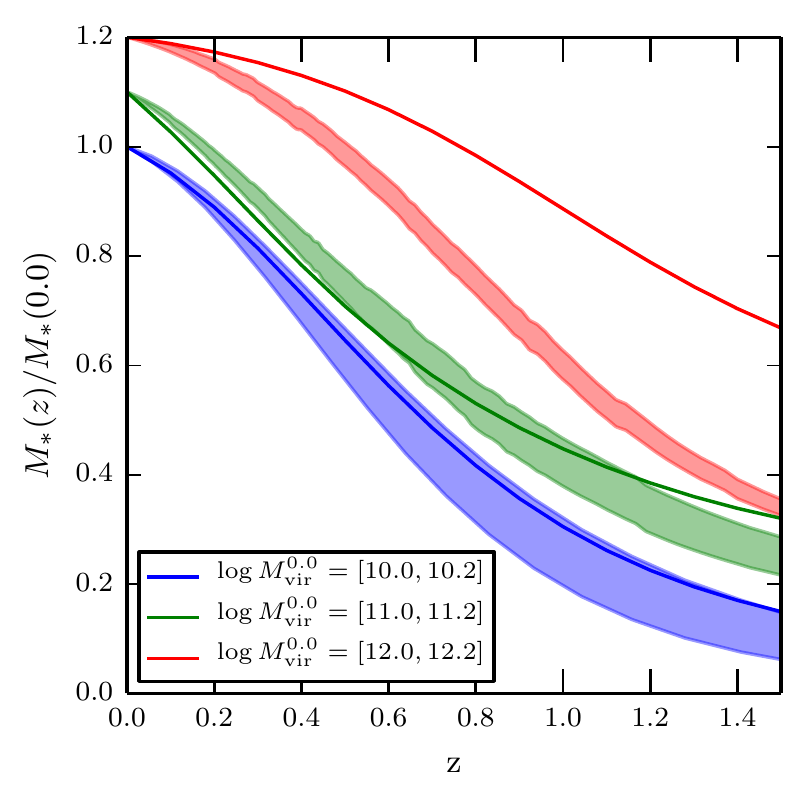}}
\caption{The growth in the stellar mass of central galaxies along the progenitor history of halos in different bins of present-day halo mass. The shaded regions show constraints inferred by \citet{behroozi_13_shmr} using detailed models which follow galaxy populations in dark matter halo merger trees from numerical simulations. We convert the $M_*-M_{\rm vir}$ relations they infer at different redshifts to $M_*-\mfrs(z)$, and show the redshift evolution of $\mstar$ at fixed $\mfrs$ using solid lines. For clarity, we shift both the solid line and the shaded region for the mass bins $\log \mvir^{0.0} = [11.0,11.2]$ and $[12.0,12.2]$ vertically by $0.1$ and $0.2$, respectively.
}
\label{fig:beh_growth_mform}
\end{figure}

For galaxy-sized halos, the slow evolution of the inner regions can be exploited to identify progenitor halos across redshifts at $z \lesssim 1$. In particular, Figure \ref{fig:mvirmform} shows that $\mfrs$ for halos of $\mvir \lesssim 10^{12} \msun$ evolves very little at $z<1$. This suggests that changes in galaxy properties at fixed $\mfrs$ can be used to deduce the evolution of these properties along their progenitor histories.

As an example, we estimate the stellar mass growth in low-mass halos at $z < 1$. For this purpose, we assume the stellar mass-halo mass relation (SHMR) and its evolution as parameterized by \citet{behroozi_13_shmr}, and convert $\mvir(z)$ to $\mfrs(z)$ to obtain the $\mstar-\mfrs$ relation as a function of redshift. We can now read off the star formation history at fixed $\mfrs$, shown as solid lines in Figure \ref{fig:beh_growth_mform}. The shaded areas show the observed star formation histories as derived from the sophisticated modeling of \citet{behroozi_13_shmr}. They convert observational data on the redshift dependence of the stellar mass function and of the specific star formation rate stellar mass relation using halo merger trees, to infer the growth histories of stellar mass in halos. On the other hand, our simple inference at fixed $\mfrs$, only utilizes constraints on the stellar mass$--$halo mass relation at different redshifts, such as the ones that can be obtained using abundance matching. The stellar mass growth histories agree well with the detailed modelling results at low halo masses, and starts to diverge once the evolution of $\mfrs$ becomes significant. Thus inferences related to the stellar mass growth histories in studies of the evolution of the galaxy--halo connection \citep[e.g.,][]{leauthaud_12_shmr} at low halo masses can be simplified by casting those results as a function of $\mfrs$. However, as we will discuss later, tying the growth histories of galaxies to the baryon accretion rate in the central regions is considerably more complicated.

\subsection{The Pseudo-evolution of Halo Mass in the Standard Definitions}
\label{sec:disc:pe}

\begin{figure} 
\centering{
\includegraphics[trim = 1mm 2mm 2mm 0mm, clip, scale=1.0]{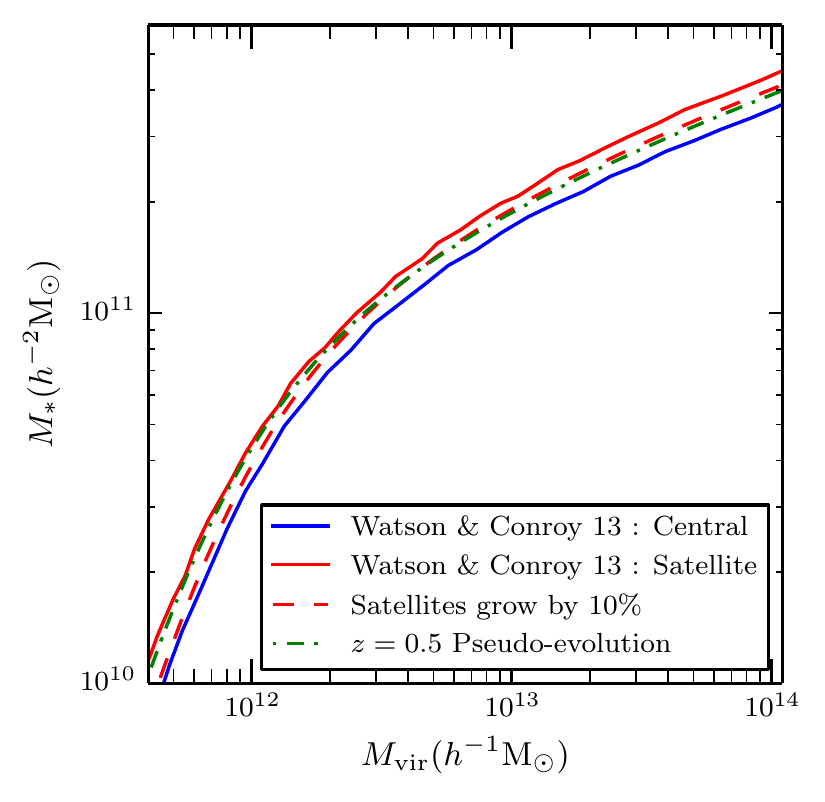}}
\caption{SHMRs for central (solid blue) and satellite (solid red) galaxies, as obtained by \citet{Watson_13_sat_shmr}. The dashed red line corresponds to their model which posits that the stellar mass in satellites grows by 10\% compared to centrals after they fall into the host halo. The dashed green curve shows the SHMR of centrals, but with halo masses measured at the average infall redshift, $z = 0.5$. Due to pseudo-evolution, the measured halo mass decreases at higher $z$, explaining most of the difference between the central and satellite SHMRs.}
\label{fig:watson}
\end{figure}

As we saw in the previous section, most of matter accreted by halos is deposited at the outskirts, while their inner regions can be relatively quiet. This finding relates directly to the mass accretion rate in different mass definitions, and to the question whether this accretion is physical or due to pseudo-evolution. 

First, we note that the halo masses in the standard definitions are within a factor of $\sim 1.5$ of $\msp$ at all $z$ (Figure \ref{fig:mvirmsp}), though higher density contrasts (such as $\mfoc$) would lead to smaller radii and thus larger differences with $\msp$. The mass accretion rate of $\msp$, on the other hand, differs significantly from the spherical overdensity definitions, and can be up to three times {\it larger} than the change in $\mtoc$ (Figure \ref{fig:dM200cdt}). This may seem surprising, since the mass growth in conventional definitions already suffers from pseudo-evolution, i.e. a growth {\it in addition} to the physical growth within the corresponding halo boundary. 

However, there are two competing differences between $d\msp/dt$ and $d\mtoc/dt$. While $d\mtoc/dt$ is larger than the physical change of the mass within $\rtoc$ due to pseudo-evolution, it also {\it misses} mass that is added outside $\rtoc$ between two epochs, i.e. the physical growth in the outskirts discussed in Section \ref{sec:disc:quiet}. While it is not a priori clear which of the two effects should dominate, Figure \ref{fig:dM200cdt} demonstrates that $d\msp/dt$ is larger than $d\mtoc/dt$ at all masses and redshifts shown, meaning that physical evolution in the outskirts dominates over the pseudo-evolution of $\mtoc$. However, for halos which are truly non-accreting ($d\msp/dt\approx 0$), pseudo-evolution would still lead to a non-zero $d\mtoc/dt$. 

Whether the evolution in $\mdelta$ under- or overestimates the true accretion rate, it does not always reflect the physical addition of matter. The relatively small difference between $\mdelta$ and $\msp$ does not reflect the physical correctness of $\mdelta$, but rather the shallow mass profile in the outer regions of halos which means that large changes in radius lead to small changes in enclosed mass.

The importance of taking pseudo-evolution into account when interpreting changes in halo mass can be illustrated by the results of \citet{Watson_13_sat_shmr} who tested the common assumption that the same stellar mass-halo mass relation (SHMR) is valid for both host and satellite halos. They showed that the relations are somewhat different and interpreted the difference as a 10\% growth in the stellar mass of satellite galaxies since the time of infall, possibly due to residual star formation before final quenching (Figure~\ref{fig:watson}). They estimated the average time of infall to be around $4$ Gyr ago, or $z \approx 0.5$. However, in their analysis (similar to all abundance matching studies) the halo masses of centrals are defined at $z=0$ while those of satellites are defined at their infall epoch ($z=0.5$). Therefore, some of the difference in the host and satellite SHMRs will arise from the pseudo-evolution of mass between these redshifts. We estimated this effect and show it with the green dot-dashed line in Figure~\ref{fig:watson}. Simply accounting for pseudo-evolution explains most of the difference in the SHMRs found by \citet{Watson_13_sat_shmr}, and does so better than the suggested blanket 10\% increase in the stellar mass of satellites across all halo masses. This agreement suggests that satellite galaxies maintain levels of star formation similar to their central counterparts for a substantial period after infall. This increase in stellar mass would depend on a satellite's halo mass, a qualitatively different view from that presented by \citet{Watson_13_sat_shmr}.

\subsection{The Accretion of Mass onto Low-mass Halos}
\label{sec:disc:lowmass}

A number of recent studies \citep{prada_06, diemand_vl_2007, cuesta_2008, diemer_13_pe, zemp_14, diemer_14_pro} have pointed out that the inner density profiles of low-mass, galaxy-sized halos evolve very little at $z\lesssim 1$. The near-constant density profiles were sometimes interpreted as a lack of new mass accretion in these halos. As we showed in the previous section, $\msp$ does, on average, increase rapidly for halos of all masses, including dwarf-sized halos, {\it implying the continued accretion of mass in all halos to low $z$.} This newly accreted mass may briefly pass through the inner regions of halos, but is primarily deposited in the halo outskirts, with no substantial change in the inner regions.

In Section \ref{sec:disc:quiet}, we explained why the newly accreted mass contributes little to the inner density profile, but this argument applies only to collisionless dark matter. Baryons, on the other hand, experience additional pressure forces and possibly dissipation and interaction with feedback-driven winds \citep{nelson_etal15}, so that their radial distribution and fate can be quite different \citep{faucher_giguere_etal11, vandevoort_etal11, vandevoort_schaye12, wetzel_nagai14, feldmann_15}. We note that the baryon accretion rate in the inner regions of halos could track the growth rate of halos (e.g., determined by $\msp$) if the trajectory of recently accreted material brings it closer to the central density peak. The accreted baryons could interact with intra-halo gas and feedback-driven winds and stay in the inner regions of halos. Such effects could conceivably explain the large baryon accretion rate and its consistency with $f_{\rm b}d\mtom/dt$, as found by \citet{wetzel_nagai14}. However, \citet{woods_14} tracked particles in their hydrodynamical simulations and showed that the late time star formation in their simulated galaxies is a result of gas cooling from a reservoir which was accreted $2-8$ Gyr earlier. \citet{nelson_etal15}, on the other hand, find that feedback-driven winds increase the time it takes the gas to reach the disk after it crosses $\rvir$ by a factor of several, but does not introduce any significant mass dependence in the accretion rate. 

Although the fate of baryons is not completely clear, our results do show that galaxy-sized systems do have a continuing supply of fresh matter to $z=0$. 

\subsection{A Halo's ``Zone of Influence'' and the Extent of the One-halo Term}
\label{sec:disc:zone}

\begin{figure} 
\centering{
\includegraphics[trim = 1mm 2mm 2mm 0mm, clip, scale=1.0]{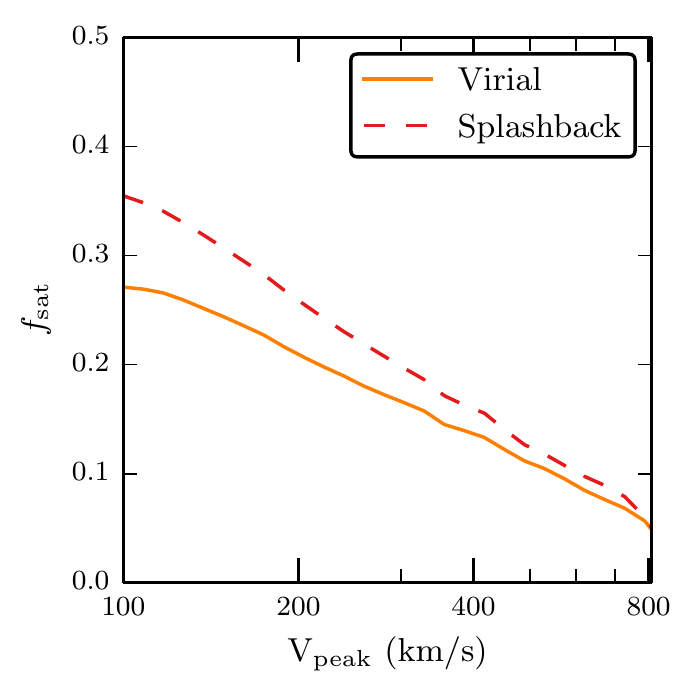}}
\caption{ Comparison of the satellite fraction as a function of the peak
        circular velocity, $\vpeak$, for subhalos identified as density peaks
        within the virial radius (solid line) to those identified within the splashback
        radius (dashed line) of more massive halos. The satellite fraction significantly
        increases for low values of $\vpeak$.
} 
\label{fig:fsat}
\end{figure}

We have shown that the splashback radius for galaxy-sized halos at low $z$  can be more than a factor of two larger than $\rtoc$. For example, for Milky Way-sized halos, the splashback radius can be as large as $600$ kpc, or two thirds of the distance to M31 (see, e.g., Figure 2 of \citealt{diemand_kuhlen08} and Figure \ref{fig:mvirmsp} above). The large $\rsp$ may explain the presence of dwarf spheroidal galaxies, such as Cetus \citep[e.g.,][]{fraternali_etal09}, at large distances, and highlights that the ``zone of influence'' of halos (i.e., their associated environmental effects) can extend to and beyond $2 \rtoc$. 

This effect was discussed previously in the context of ``backsplash satellites'' which were found in substantial numbers beyond the halo radii in the standard definitions \citep[][see also \citealt{wetzel_etal14} for in depth analysis of this issue and more recent references]{balogh_etal00,mamon_etal04,gill_etal05}. Their occurrence at large radii is not surprising because satellite subhalos effectively trace the distribution of accreted matter. This also means that the radial number density profiles of subhalos in simulations and galaxies in observed clusters may exhibit the steepening at the splashback radius (see further discussion in Section \ref{sec:disc:detections}). Considering the extent of the splashback radius, it is unnecessary to invoke special ejection processes to explain the presence of subhalos that have orbited their host at large radii and are deficient in gas or show suppressed star formation \citep[e.g.,][]{balogh_etal00, solanes_etal02, geha_etal12, wetzel_etal12}. Rather, these galaxies may reflect a population of halos that accreted onto their host halo late and are located close to the apocenter of their orbit at $\sim\rsp$. 
 
As $\rsp$ corresponds to the physical boundary between the inner and infall regions of halos, $\rsp$ may also be a natural boundary of the one-halo term in halo models of structure formation. Using such a boundary would result in the reclassification of some isolated halos as subhalos. In Figure \ref{fig:fsat}, we explore the magnitude of this effect by comparing the fraction of density peaks which are subhalos (i.e., the satellite fraction) as a function of their peak circular velocity, $\vpeak$, in the standard virial definition to that in the splashback case\footnote{We have calculated the accretion rate $\Gamma$ for individual halos in the Bolshoi simulation and assumed zero scatter between $\Gamma-\rsp$ in order to define the boundaries of these halos. Subhalo classifications were then revised based on their updated boundaries.}. For low values of $\vpeak\sim100\kms$, the satellite fraction based on $\rsp$ shows an increase of $30$ percent. 

The reclassification of isolated halos also has implications for assembly bias. It is well-known that the large-scale bias of halos depends upon their assembly history \citep{Gao:2005, Wechsler:2006, Li:2008, Dalal:2008}. At the low (high) mass end, the large scale bias of halos is (anti-) correlated with their concentrations. The low-mass halos that are classified as isolated in the virial definition but as subhalos in the splashback definition are not a random subsample of the parent population. They tend to be in the vicinity of more massive halos, have higher concentrations, and higher bias values. Thus, their reclassification as satellites can reduce the assembly bias of low-mass halos to some extent, but it does not explain the bulk of the effect \citep[see also][]{wang_etal09, Ludlow:2009}.

In halo models, accurate modeling of the transition region between the one- and two-halo terms has remained a challenge. The use of $\rsp$ as the boundary in such models may lead to some progress, but requires an accurate calibration of the halo mass function and bias as a function of $\msp$ using cosmological simulations. We could convert existing mass function and bias formulae for $\mtom$ \citep[e.g.,][]{tinker_etal08, tinker_etal10} using the conversions in Equations \ref{eq:rsp_nu} and \ref{eq:msp_nu}, but given the large scatter in those relations the results would be approximate. Instead, we will require robust methods to measure $\rsp$ and $\msp$ for {\it individual} halos in simulations, rather than using the mean or median density profiles of halo samples.

\subsection{Possible Observational Detections of the Splashback Radius}
\label{sec:disc:detections}

Interestingly, the steepening of the density profile associated with the splashback radius may already have been detected in several galaxy clusters. \citet{rines_etal13} used the velocity caustic method to derive mass profiles of a number of clusters to large radii. Figure 11 in their paper shows that the radial density profiles of massive clusters sharply transition to slopes steeper than the value of $-3$ expected from the NFW profile. The steepening occurs at radii where $\rho(r)\sim 10-20\rho_{\rm m}$, roughly where the splashback radius is expected to be located for typical halos with $\Gamma\sim 1-2$ \citep[see Figure \ref{fig:profiles} above and  Figure 2 in][]{diemer_14_pro}. 

Furthermore, a deep {\sl Chandra} observation of Abell 133 (Vikhlinin et al., in preparation) shows a strong steepening of the gas density profile just beyond $\rtoc$. Although the steepening is observed in gas rather than in the total mass profile, hydrodynamic simulations show that the gas profile exhibits a steepening at the same radius as the dark matter profile \citep{lau_etal14}. According to the calibration of Equation \ref{eq:rsp_gamma}, the proximity of the splashback radius to $\rtoc$ would correspond to a very high mass accretion rate onto this relatively relaxed, ``cool core'' cluster. In the future, the density profile steepening associated with splashback may be independently verified with mass profile measurements using weak lensing or via radial profiles of number density of galaxies. 

Finally, \citet{tully15} recently argued that a radius qualitatively similar to the splashback radius which the author called ``the second turnaround radius'' should be used to define the halo extent in analyses of observational samples. \citet{tully15} presents evidence for a sharp drop in the number density and velocity dispersion profiles of galaxies in a number of systems from cluster-sized to the Local Group.  We note that for the Milky Way and M31, \citet{tully15} derives a splashback radius of $\approx 290$ kpc, considerably smaller than what we would expect for systems of this mass on average: $\rsp \approx 1.4-1.5\rtom\approx 400-600$ kpc (see Figure \ref{fig:nu_rsp}). Overall, \citet{tully15} find a scaling of $\rsp \approx 1.33 \rtoc$, significantly smaller than our calibration. This can be traced to some of the simplifying assumptions made in his calculation of the radius. 

\subsection{How to Estimate $\mfrs$ and $\msp$ in Practice?}
\label{sec:disc:practice}

In the standard spherical overdensity definition, the radius $\rdelta$ enclosing a given overdensity $\Delta$ depends on the amount of mass within $\rdelta$, but not on the {\it distribution} of this mass. In contrast, estimating $\mfrs$ demands knowledge of the scale radius (or concentration; Equation \ref{eq:m4rs}), and estimating $\msp$ requires a measurement of the density profile at the outermost radii of a halo, or knowledge of its accretion rate (Equation \ref{eq:rsp_gamma}). In simulations, computing $\mfrs$ presents no problem, since the scale radius is routinely computed by halo finders. The splashback radius can be found by finding the radius of steepest slope in halo density profiles, or by fitting those profiles with the fitting function of \citet{diemer_14_pro}. It may be possible to develop more robust methods not relying on spherically averaged profiles. For example, a clearer picture of the density caustic emerges from the distribution of halo particles on the phase-space plane ($v_{\rm r}-r$, where $v_{\rm r}$ is the radial velocity with respect to the halo center; \citealt{adhikari_14}). 

Observationally, measurements of $\rs$, $\rsp$, or the accretion rate are challenging, but $\mfrs$ and $\msp$ can be estimated using their average relations with conventionally defined masses. For $\mfrs$, all we need is a well-calibrated model for halo concentrations to estimate $\rs$ \citep[e.g.,][]{diemer_14_cnu}. For $\rsp$ and $\msp$, we have quantified $\rsp$ and $\msp$ as a function of $\rtom$ and $\mtom$ and the mass accretion rate, given in Equations (\ref{eq:rsp_gamma}) and (\ref{eq:msp_gamma}). For the case where the mass accretion rate of a halo is not known, we have given $\rsp$ and $\msp$ as a function of peak height (or mass) in Equations (\ref{eq:rsp_nu}) and (\ref{eq:msp_nu}). However, we note that there is one easily accessible quantity that contains information beyond halo mass: the concentration. Since the concentration is intimately related to the mass accretion history of a halo \citep[e.g.,][]{ludlow_13_mah}, we expect there to be a relation between $\Gamma$ and concentration. We find that, in principle, this relation depends on redshift and mass in a non-trivial manner. However, at $z = 0$ we find a simple, mass-independent relation,
\begin{equation}
\Gamma \approx 3.43 - 2.74 \log_{10}(c_{\rm vir}) \,.
\label{eq:gamma_c}
\end{equation}
We note that this relation was only calibrated for the cosmology used in this paper, and for our particular definition of $\Gamma$. Nevertheless, the relation can be used to estimate the mass accretion rate, and thus $\rsp$ and $\msp$, at $z = 0$. 

For convenience, we have implemented all the calibrations of $\rsp$, $\msp$, and $\mfrs$ given in this paper, as a function of mass accretion rate, peak height, and mass, in the public python code {\tt Colossus}. This stand-alone module can be downloaded at {\tt www.benediktdiemer.com/code}.


\begin{figure*} \centering{
\includegraphics[trim = 2mm 1mm 2mm 2mm, clip, scale=1.02]{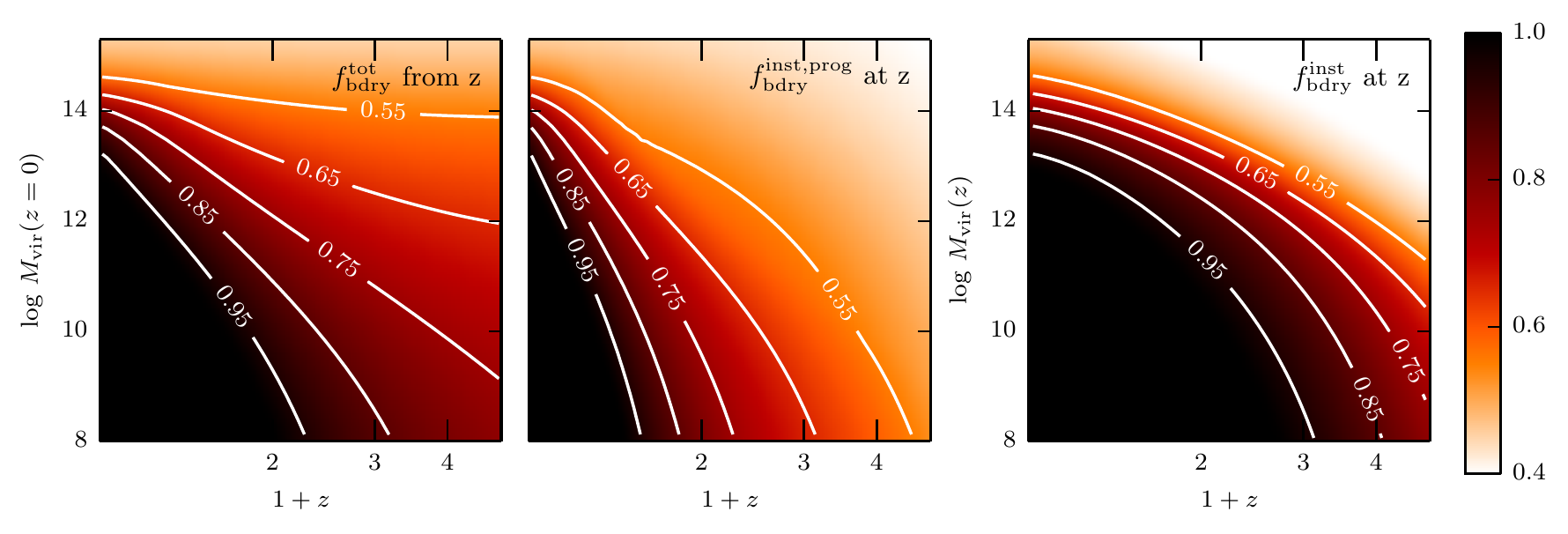}}
\caption{
Fractional contribution of pseudo-evolution to halo mass growth, $\fpe$. {\it Left panel:} The fraction of mass growth due to the shift in the boundary term, $\fpe^{\rm tot}$, between redshift $z$ and $z = 0$ for halos of different virial masses at $z=0$ (Equation (\ref{eq:pe})). {\it Center panel:} Same as the left panel, but showing the instantaneous fraction at redshift $z$ rather than the overall fraction since $z$ (Equation (\ref{eq:pediff})). {\it Right panel:} Same as the center panel, but as a function of $\mvir$ at redshift $z$ rather than as a function of the descendant mass.
}
\label{fig:pefrac}
\end{figure*}

\section{Conclusions}
\label{sec:conclusion}

We have closely examined the definition of the boundary of CDM halos and the associated mass. We have argued that the most natural definition of physical halo boundary is the splashback radius corresponding to the apocenters of orbits of the most recently accreted matter. As shown by \citet[][see also \citealt{adhikari_14}]{diemer_14_pro}, this splashback radius manifests itself as a sharp steepening of the outer density profiles of halos (illustrated in Figures \ref{fig:viz} and \ref{fig:profiles}). In this paper, we calibrate the dependence of the splashback radius on the halo mass accretion rate and mass (characterized by the peak height $\nu$). We present a detailed comparison of the evolution of the splashback radius and mass to the evolution of radii and masses in the commonly used spherical overdensity definitions. In the Appendix we present a detailed analysis of the components that contribute to the growth of halo mass. Our main findings and conclusions are summarized below.

\begin{enumerate}

\item The splashback radius depends primarily on the mass accretion rate of halos, with some cosmological dependence on the mean matter density of the universe, $\Omega_{\rm m}(z)$  (Equations (\ref{eq:rsp_gamma}) and (\ref{eq:msp_gamma})). For halos rapidly accreting mass, $\rsp\lesssim \rtom$, while for slowly accreting halos $\rsp$ can be as large as $\sim 1.5-1.6\rtom$, up to $\approx 2 \rtoc$. Thus, a halo and its environmental effects may extend well beyond the commonly used ``virial'' radii. 

\item A comparison of the growth rates of $\msp$ and masses in the standard definitions $\mtoc$, $\mvir$ and $\mtom$ shows that, at low $z$, $\msp$ increases {\it faster} than $\mtoc$, $\mvir$, and $\mtom$, indicating a substantial physical mass growth at low $z$. Even the smallest halos continue to accrete new matter which may fuel their star formation.

\item To characterize the mass evolution in the inner regions of halos, we introduce a second new mass definition, $\mfrs$, the mass enclosed within $4$ scale radii of a halo. This mass is manifestly not affected by pseudo-evolution because it is not tied to any reference density. In the fast mass growth regime, $\mfrs \approx \mvir\approx \msp$, while in the slow mass growth regime $\mfrs$ approaches a constant. In the latter regime, the interpretation of the evolution of the $\mstar-\mfrs$ relation thus becomes very simple: at fixed $\mfrs$, the evolution of the relation represents the change in the stellar mass of a given halo along its main progenitor branch. We derive the star formation histories of low-mass galaxies in this manner, and show that $d\mstar / dt(\mfrs = {\rm const})$ is in good agreement with SFHs inferred from more complex modeling. 

\item A comparison of the evolution of $\mfrs$ and $\msp$ shows that the mass within the inner regions of many halos evolves slowly, even though the halo experiences substantial mass growth overall. The newly accreted mass is deposited with a relatively shallow radial density profile and thus contributes most significantly to the outer regions. The co-existence of a relatively slow evolution of the interior mass distribution with an active accretion in the outskirts means that systems that appear quiescent may actually actively grow, particularly because much of the mass growth is due to the accretion of diffuse mass and small halos rather than due to major mergers. Thus, in cluster-sized halos, relaxed ``cool-core'' inner regions can co-exist with active accretion in the outer regions.

\item We showed that the relation between the mass accretion rates of $\msp$ and spherical overdensity definitions, $\mdelta$, is complicated by the pseudo-evolution affecting the latter. We illustrated the importance of accounting for pseudo-evolution by showing that it can accurately explain the differences in the central and satellite $\mstar-\mvir$ relations found by \citet{Watson_13_sat_shmr}, if halos that are satellites at $z = 0$ were, on average, accreted at $z = 0.5$. After accounting for pseudo-evolution, the data imply an even larger amount of post-accretion star formation in satellite halos than the 10\% increase of stellar mass inferred by \citet{Watson_13_sat_shmr}.
\end{enumerate}

Finally, we alluded to possible observational detections of the splashback radius in galaxy clusters. A better observational characterization of the halo environment close to the splashback radius will be an interesting avenue to pursue in the near future. Likewise, it will be interesting to explore the implications of a larger halo extent of slowly accreting halos for the statistics of their satellite population, the halo mass function, bias, and the halo model of galaxy clustering. In particular, we plan to explore whether statistics such as the halo mass function are more universal when recast as a function of $\msp$. 


\vspace{0.5cm}

We are grateful to Susmita Adhikari and Neal Dalal for enlightening discussions
on the splashback radius and for sharing the draft of their paper prior to
publication. We thank Philip Mansfield for his help with the production of the
density maps shown in Figure 3.  We would also like to thank Matt Becker, Peter
Behroozi, Kevin Bundy, Joanne Cohn, Robert Feldmann, Phil Hopkins, Alexie
Leauthaud, Masahiro Takada, Andrew Wetzel, Martin White, and Freeke van de Voort for useful
discussions.  SM is supported by Kavli IPMU, a World Premier International
Research Center Initiative (WPI Initiative), MEXT, Japan, by the FIRST program
``Subaru Measurements of Images and Redshifts (SuMIRe)'' of the Council for
Science and Technology Policy, Japan. This work was also supported by NASA ATP
grant NNH12ZDA001N and by the Kavli Institute for Cosmological Physics (KICP) at
the University of Chicago through grants NSF PHY-0551142 and PHY-1125897.  Both
Kavli IPMU and KICP are supported through a generous endowment from the Kavli
Foundation and its founder Fred Kavli. This work was also supported in part by
the National Science Foundation under Grant No. PHYS-1066293 and the hospitality
of the Aspen Center for Physics. This work was completed in part using the computing resources provided by the University of Chicago Research Computing Center.


\appendix

\section{Components of Halo Mass Growth}
\label{sec:toy}

Here we consider in detail the factors that contribute to the evolution of the standard mass definitions employing the spherical radius enclosing a given density contrast \citep[see also detailed discussion in Section 2 of][]{diemer_13_pe}. We quantify the contribution of mass growth due to the change of halo boundary radius for a large range of halo masses and redshifts, and discuss how this contribution should be interpreted in the context of halo mass growth.

Let us consider the change in mass, $\mdelta$, of a spherical overdensity halo between two redshifts $z_\rmi$ and $z_\rmf$. As noted in \citet{diemer_13_pe}, in general this change will consist of two components: 
\begin{eqnarray}
M(z_\rmf)&=&M(z_\rmi)+ \Delta M_{\rm tot}(z_\rmi,z_\rmf) \nonumber \\ 
         &=&M(z_\rmi)+\Delta M_{\rm bdry}(z_\rmi,z_\rmf) + \Delta M_{\rm acc}(z_\rmi,z_\rmf)\,,
\label{eq:delm}
\end{eqnarray}
where
\begin{equation}
\Delta M_{\rm acc}(z_\rmi,z_\rmf) = \int_{z_\rmi}^{z_\rmf} dz \int_{0}^{\rdelta(z)} dr\,4\pi r^2
\frac{ d\rho(r,z)}{d z}
\label{eq:mphys}
\end{equation}
is due to the actual change of the density profile {\it within} the halo boundary, while 
\begin{equation}
\Delta M_{\rm bdry}(z_\rmi,z_\rmf) = \int_{\rdelta(z_\rmf)}^{\rdelta(z_\rmi)} dr\,4\pi r^2
\rho(r,z_\rmc)\,.
\label{eq:mpe}
\end{equation}
is the mass change that arises solely because the boundary of the halo increases due to decreasing reference density. We have explicitly retained the redshift dependence of the halo radius $\rdelta$, so that the integral in Equation (\ref{eq:mpe}) contains the density at the position $r$ and redshift $z_\rmc$ where $\rdelta(z_\rmc) = r$.
We can further expand this term,
\begin{eqnarray}
\label{eq:mpeexp}
\Delta M_{\rm bdry} (z_\rmf) &=& \int_{R(z_\rmi)}^{\tilde{R}(z_\rmf)} \rmd r \,4 \pi r^2 \,\rho(r, z_\rmi) \nonumber \\ 
                         &&+ \int_{R(z_\rmi)}^{\tilde{R}(z_\rmf)} \rmd r \,4 \pi r^2 \,[\rho(r, z_\rmc)-\rho(r, z_\rmi)] \nonumber \\
                         &&+ \int_{\tilde{R}(z_\rmf)}^{R(z_\rmf)} \rmd r \,4 \pi r^2 \,\rho(r, z_\rmc) \,.
\end{eqnarray}
The first integral represents the mass change that would arise due to the expanding halo boundary if the density profile at $z_{\rmi}$ were to remain unchanged (static), while the other two terms correspond to a changing density profile at intermediate redshifts $z_\rmc$ and would be zero for a static profile. Note that we integrate the first term only out to $\tilde{R}(z_\rmf)$ which denotes the boundary of the halo at $z_\rmf$, inferred from the density profile at redshift $z_\rmi$.  

The mass change $\Delta M_{\rm acc}$ clearly corresponds to the newly accreted mass within the radius $\rdelta$. The $\Delta M_{\rm bdry}$ component, however, in general contains both mass that was accreted during the interval $\Delta z=z_{\rmi}-z_{\rmf}$ and mass that was accreted before, at $z>z_{\rmi}$, but located at $r>\rdelta(z_{\rmi})$. The relative fractions in these contributions to the mass growth will depend on the actual choice of the value of $\Delta$ used to define the halo boundary and the evolutionary stage of a halo. 

Let us consider the fraction of mass change due to the evolution of the halo boundary:
\begin{equation}
f_{\rm bdry}^{\rm tot}(M[z_\rmf],z_\rmf,z_\rmi)=\left. \Delta M_{\rm bdry}\right/\Delta
M_{\rm tot} \,,
\label{eq:pe}
\end{equation}
This fraction is shown in the left panel of Figure \ref{fig:pefrac} for $z_\rmf=0$ as a function of $z_\rmi$ and halo mass. For a halo of mass $M(z_\rmf)$ identified at redshift $z_\rmf$, the fraction of the instantaneous mass growth in its main progenitor at redshift $z$ due to the change in its boundary is given by
\begin{equation}
f_{\rm bdry}^{\rm inst,prog}(M[z_\rmf],z_\rmf,z) = \left.\left( \frac{d \Delta M_{\rm
bdry}}{dz} \right)_z \,\right/\,
\left(\frac{d \Delta M_{\rm tot}}{dz}\right)_z\,.
\label{eq:pediff}
\end{equation}
This fraction is shown in the middle panel of Figure~\ref{fig:pefrac}.

Finally,
the instantaneous fraction of mass growth due to the boundary change for a halo of
mass $M$ at redshift $z_\rmf$ is obtained by taking the limit $z \to z_\rmf$
of the above equation,
\begin{equation}
f_{\rm bdry}^{\rm inst}(M,z_\rmf) = \lim_{z\to z_\rmf} f_{\rm bdry}^{\rm inst,prog}(M[z_\rmf],z_\rmf,z)
\label{eq:pediffz}
\end{equation}
and is shown in the right panel of Figure~\ref{fig:pefrac}. The left and central panels can be used to estimate the contribution of the boundary evolution to the growth of halo mass along the main progenitor branch of a $z=0$ halo. The fractions displayed in Figure \ref{fig:pefrac} are computed using the mass accretion history and concentration model of \citet{zhao_09_mah}. We compute the density profile of the main progenitor, $\rho(r,z)$, assuming the Navarro-Frenk-White density profile \citep[][hereafter NFW]{navarro_97_nfw}.

The left panel of Figure \ref{fig:pefrac} shows that between $z=3$ and $z=0$ the mass change due to the boundary evolution accounts for more than half of the change in halo mass, even for massive $10^{14}\msunh$ halos. This fraction tends to unity for low-mass halos at lower redshifts. Note, in particular, the large values of $f_{\rm bdry}^{\rm inst}$  for galaxy-sized halos at $z\lesssim 1$. Moreover, $\gtrsim 85\%$ of mass growth for halos of $M_{\rm vir}\lesssim 10^9\ \rm M_{\odot}$, expected to host dwarf galaxies, at $z\lesssim 2-3$ is due to the halo boundary change.  

However, as we discussed in Section \ref{sec:peakcollapse}, the mass increase within the splashback radius $\rsp$ between two epochs $z_{\rmi}$ and $z_{\rmf}$ using Equations (\ref{eq:delm})--(\ref{eq:mpe}) is due to the {\it new} mass shells that have entered the splashback radius in the interval $\Delta z=z_{\rmi}-z_{\rmf}$. However, $\Delta M_{\rm bdry}$ in Equation (\ref{eq:mpe}) is positive and actually represents a significant fraction of the total mass increase, $\Delta M_{\rm tot}$. This illustrates that in the presence of actual mass accretion, $f_{\rm bdry}=\Delta M_{\rm bdry}/\Delta M_{\rm tot}>0$ should not be of particular concern. In fact, for the choice of $\rsp$ as the halo boundary, the entire $\Delta M_{\rm bdry}$ should correspond to the accretion of new mass during the interval $\Delta z$, and the change of mass within $\rsp$ {\it should not have any contribution from pseudo-evolution.} 
For halo mass definitions using boundaries $R<\rsp$, a fraction of $f_{\rm bdry}$ will correspond to pseudo-evolution rather than to new accretion of mass with contribution of PE increasing with decreasing physical mass accretion rate. 

In summary, Figure~\ref{fig:pefrac} demonstrates that the mass growth due to the evolution of the halo boundary is important for all halo masses of astrophysical interest, and over a wide range of redshifts. How this change should be interpreted, however, depends on the evolutionary stage of a halo and on the specific choice of $\Delta$. 


\bibliographystyle{apj}
\bibliography{sf}

\end{document}